\documentclass[conference]{IEEEtran}
\IEEEoverridecommandlockouts
\usepackage{cite}
\usepackage{amsmath,amssymb,amsfonts}
\usepackage{algorithmic}
\usepackage{physics}
\usepackage{graphicx}
\usepackage{textcomp}
\usepackage{xcolor}
\usepackage{hyperref}
\def\BibTeX{{\rm B\kern-.05em{\sc i\kern-.025em b}\kern-.08em
    T\kern-.1667em\lower.7ex\hbox{E}\kern-.125emX}}

\usepackage{subcaption}
  
\usepackage{tikz}
\usetikzlibrary{quantikz, shapes, arrows}
\def\orctrl#1{\control[style={fill=white, draw=black, shape=forbidden sign, minimum size=5pt}]{} \vqw{#1}}
\def\norctrl#1{\control[style={fill=white, draw=black, shape=rectangle, minimum size=5pt}]{} \vqw{#1}}
\def\rzz#1{\phase{ZZ}}
\def\rxx#1{\phase{XX}}
\def\ryy#1{\phase{YY}}

\usepackage{pgfplots}
\pgfplotsset{compat=1.17}

\usepackage{verbatimbox}

\begin{document}

\title{QPack Scores: Quantitative performance metrics for application-oriented quantum computer benchmarking\\
}

\author{
\IEEEauthorblockN{Huub Donkers}
\IEEEauthorblockA{\small \textit{Quantum \& Computer Engineering} \\
\textit{Delft University of Technology}\\
Delft, The Netherlands \\
h.j.donkers@student.tudelft.nl}
\and
\IEEEauthorblockN{Koen Mesman}
\IEEEauthorblockA{\small \textit{QBlox} \\
Delft, The Netherlands \\
kmesman@qblox.com}
\and
\IEEEauthorblockN{Zaid Al-Ars}
\IEEEauthorblockA{\small \textit{Quantum \& Computer Engineering} \\
\textit{Delft University of Technology}\\
Delft, The Netherlands \\
z.al-ars@tudelft.nl}
\and
\IEEEauthorblockN{Matthias Möller}
\IEEEauthorblockA{\small \textit{Numerical Analysis, DIAM} \\
\textit{Delft University of Technology}\\
Delft, The Netherlands \\
m.moller@tudelft.nl}
}

\maketitle

\thispagestyle{plain}
\pagestyle{plain}

\begin{abstract}
This paper presents the benchmark score definitions of QPack, an application-oriented cross-platform benchmarking suite for quantum computers and simulators, which makes use of scalable Quantum Approximate Optimization Algorithm and Variational Quantum Eigensolver applications. Using a varied set of benchmark applications, an insight of how well a quantum computer or its simulator performs on a general NISQ-era application can be quantitatively made. This paper presents what quantum execution data can be collected and transformed into benchmark scores for application-oriented quantum benchmarking. Definitions are given for an overall benchmark score, as well as sub-scores based on runtime, accuracy, scalability and capacity performance. Using these scores, a comparison is made between various quantum computer simulators, running both locally and on vendors' remote cloud services. We also use the QPack benchmark to collect a small set of quantum execution data of the IBMQ Nairobi quantum processor. The goal of the QPack benchmark scores is to give a holistic insight into quantum performance and the ability to make easy and quick comparisons between different quantum computers.\\
\end{abstract}


\section{Introduction} \label{sec:intro}



Over the past two decades, quantum computers have evolved rapidly. Since the realization of the first physical quantum gate by NIST in 1995~\cite{first_q_gate}, DiVincenzo's famous list of criteria to realize a quantum computer proposed in 1996~\cite{DiVincenzo_2000} and the Oxford university's implementation of a two-qubit quantum computer that solves Deutsch's problem~\cite{first_qpu}, quantum computers have achieved many milestones. In 2019, Google reported to have achieved quantum supremacy: a major milestone in quantum computing. Their 53-qubit system could perform a task in 200 seconds where today's classical supercomputers would need about 10 thousand years~\cite{supremacy}, although this claim was criticized by IBM to actually be 2.5 days~\cite{ibm_rebuttal}. Current state-of-the-art quantum computers push the technology even further, such as Google's 72-qubit Bristlecone~\cite{google_bristlecone}, Rigetti's 80-qubit Aspen-M-1~\cite{rigetti_aspenM1} and IBM's 127-qubit Eagle~\cite{ibm_eagle} quantum processors. 

Still, despite these advances, the current state of quantum computing is labeled as the Noisy Intermediate-Scale Quantum (NISQ) era~\cite{nisq}, emphasizing the fact that although quantum computing is becoming an important tool to push technological boundaries, it is still in its infancy and many challenges still need to be solved to scale up quantum computing performance. 

When scaling up quantum computers, the number of qubits is not the only important goal. Qubit decoherence time and gate fidelity play an important role in the ability to increase quantum processor size. As the number of qubits increases, noise and cross-talk tend to increase the error rate rapidly~\cite{scale_challenge}. After all, what is the benefit of a 1000 qubit device that cannot produce a meaningful output? It is thus important to benchmark this low-level noisy behavior and see where performance gains are feasible. Much work has gone into Quantum Characterization, Verification and Validation (QCVV), which targets measurements of single or dual qubit gates' noise behavior~\cite{volumetric_benchmarks}. Although important, these low-level performance characteristics do not capture the quantum computers' overall performance, giving rise to the problem that the high-level performance of a quantum computer cannot properly be understood by only evaluating its individual components. This illustrates the need for holistic quantum benchmarks that allow for the possibility to compare the performance of different hardware platforms on the system level.

In this paper, a continuation of prior work~\cite{QPack_benchmark_paper} is presented, focusing on using quantum execution data obtained from the QPack benchmark to provide benchmark scores to measure quantum performance. The goal of the QPack benchmark is the ability to make a quantitative comparison between quantum computers (both physical realizations and simulators) by scoring a quantum computer on multiple characteristics, such as quantum runtime, output accuracy and scalability. This gives insight in the areas that a quantum computer excels in and where it can improve performance. It can also provide a general idea of what type of hardware should be used for a desired application. For example, an application may require a more accurate result without runtime being a concern, and vice versa. Alternatively, some applications might benefit from a denser connectivity of qubits rather than a large number of qubits. As such, a quantum computer can be selected based on benchmark performance.

QPack is built atop the cross-platform library LibKet~\cite{libket, libket_docs}, which allows for a single compilation of a quantum circuit and execution on a variety of quantum computers and simulators. In this paper, a quantum computer is considered to take an quantum instruction set as input and returns a state distribution histogram. This covers the whole system of actual quantum hardware, control system, qubit mapping and gate scheduling, or its simulation thereof. The Quantum Approximate Optimization Algorithm (QAOA)~\cite{QAOA_fahri} and the Variational Quantum Eigensolver (VQE)~\cite{VQE} are used as NISQ-era quantum applications for which quantum performance data is collected, as these variational quantum circuits can already give meaningful results on small and noisy quantum computers. This measured data is then transformed into quantum performance sub-scores and an overall score based on these sub-scores is finally computed.

The rest of this paper is organized as follows. Section \ref{sec:q_benchmarking} provides an overview of the current state of quantum application-level benchmarks and highlights some recent benchmark proposals. In Section \ref{sec:metrics}, quantum execution data obtained during benchmarking are presented, as well as criteria for benchmark scores. This lays the foundation for the benchmark scores, which are defined in Section \ref{sec:score_def}. Their application on a selection of local and remote quantum simulators is presented in Section \ref{sec:results}, followed by results obtained on actual quantum hardware. The results are discussed in Section \ref{sec:discussion}. The paper ends with the conclusions in Section \ref{sec:conclusion}.
\section{Quantum application-oriented benchmarking} \label{sec:q_benchmarking}

\begin{table*}[h]
    \centering
    \caption{QPack quantum execution data description}
    \resizebox{\linewidth}{!}{%
    \begin{tabular}{l|l}
    Metric & Description \\ \hline \hline
    Depth & The depth of the untranspiled quantum circuit as implemented in QPack. Although the circuit needs  \\
          & to be transpiled to fit a certain qubit layout, QPack does not take this transpiled depth into \\
          & consideration when evaluating performance, as this is a job for the quantum computer under test \\ \hline
    Expectation value & Expectation value of the VQA problem with the optimized parameters \\ \hline     
    Expectation value baseline & Expectation value of the QuEST noiseless simulator. This data serves as a baseline for how well \\
                               & the classical optimizer works on a given problem set. \\ \hline
    Expectation value optimal & Theoretical best expectation value that can be achieved for a given problem and problem size \\ \hline          
    Job durations [ms] & Duration of the circuit execution time of each quantum job of an optimizer iteration. This is the time \\
                       & that the quantum computer executes the quantum circuit for a given amount of shots (default: 4096) and \\
                       & does not include circuit transpiling, optimization or other latencies.\\ \hline
    Optimal params & Optimized parameters of the VQA by the classical optimizer\\ \hline
    Optimizer durations [ms] & Time for the optimizer to finish one optimization iteration \\ \hline
    Optimizer iterations & Number of iterations the optimizer needed to find the minimal value\\ \hline
    Qubits & Number of qubits used for this problem size (circuit width) \\ \hline
    
    Total classic duration [s] & Total duration minus quantum duration. This includes optimization time, communication time, wall time \\
                               & and data read/write time. All time that is not quantum job execution. \\ \hline
    Total quantum duration [s] &  Sum of quantum job durations \\ \hline
    \end{tabular}
    }%
    \label{tab:metrics}
\end{table*}

As the growth of quantum computing hardware accelerates, so does the need for the assessment of their performance. A common way of indexing quantum performance of a quantum computer as a whole is by determining its Quantum Volume (QV)~\cite{IBM_benchmark}. Since IBMs introduction of the Quantum Volume in 2019, a wide variety of Volumetric Benchmarks have been proposed. 

The original QV metric was proposed as a single number performance metric for near-term quantum computers of modest size ($<$ 50 qubits).Quantum execution data is obtained by running randomized circuits with a width of $m$ qubits and depth of $d$. The QV is then defined as $2^m$ for the largest square circuit ($m=d$) that a quantum computer is able to successfully implement. For the QV, success means a probability of sampling heavy output above $\frac{2}{3}$. In principle, the higher a quantum computer's quantum volume, the more complex the circuit it can successfully execute. 

In February 2021, Atos Quantum Laboratories proposed a new benchmark, dubbed the Atos Q-score~\cite{Atos_benchmark}. The Q-score is defined as the maximum number of qubits that can be used effectively to solve the MaxCut combinatorial optimization problem using QAOA. The score is computed by running 100 random Erdös-Renyi graphs~\cite{ER_graph} $G(n, p = 0.5)$ and see for which $n$ the algorithm has an average energy above a certain threshold $\beta^\star$. Here, $\beta^\star$, has a value between 0 and 1, where 0 indicates a random result and 1 the result of an exact solver. Atos has arbitrarily set $\beta^\star = 0.2$, which they deemed adequate for benchmarking purposes.

Another volumetric benchmark was proposed by the Quantum Economic Development Consortium (QED-C) in October 2021~\cite{QEDC_benchmark}. This open source benchmark probes a quantum computer with small applications, such as Deutsch-Josza, Grover and VQE (see~\cite{QEDC_repo}), for which the problem sizes are varied, mapping the fidelity of the results as a function of circuit width and depth, hence making it a volumetric benchmark. Other than this primary metric, the benchmark also provides insight into runtime and the ratio between programmed and transpiled circuit depth. A remark is made that runtime metrics are currently rudimentary, as quantum providers can have different definitions of quantum execution time.

IonQ is also pursuing defining a single number performance metric for quantum computing, proposing the Algorithmic Qubits (\#AQ) in February 2022~\cite{alg_qubits}. Taking inspiration from the QED-C work, it also derives its benchmark metrics from NISQ executable quantum applications. IonQ defines $\text{\#AQ}= N$ for the largest box with $N$ qubits and a circuit depth of $N^2$ for which a quantum circuit meets a success criteria. The set of algorithms that have been selected for this benchmark are the same found in the QED-C repository~\cite{QEDC_repo}. The main difference between the QED-C and IonQ benchmarks is the definition of circuit depth (both single and two-qubit gates vs only two-qubit gates, respectively) and the success criteria of a quantum circuit. 

Alternative approaches to quantum benchmarking rather than using quantum volume were proposed by QuSoft in 2021 and Super.tech in 2022. Both propose a hardware-agnostic benchmark using real-world NISQ applications. QuSoft mainly emphasizes the interplay between various relevant characteristics of quantum computers, such as qubit count, connectivity, and gate and measurement fidelity~\cite{qusoft}. Super.tech focuses on applications that are not only chosen based on their real-world purpose, but also their coverage of the application space, described as feature vectors~\cite{supermarq}. In contrast to a single number of QV, these benchmarks define a different score for each quantum application. Thus, the benchmark does not provide a single score, but gives a rather insightful way to see what type of problems a quantum computer excels at and which it does not. These benchmark scores mainly evaluate the noisy behavior of quantum computers.

Stepping away from measuring performance of the output state of a quantum computer, IBM introduces yet another performance metric based on the execution time in October 2021~\cite{clops}: the CLOPS metric (Circuit Layer Operations Per Second) measures how many QV circuits a quantum computer can execute per time unit. This benchmarking approach may give more insight into how well quantum computers perform in contrast to classical computers and may identify where performance improvements with respect to runtime might be gained. 

QPack proposes a single number benchmark score composed of four sub-scores which is based on the evaluation of different Variational Quantum Algorithm (VQA) applications with a hardware-agnostic approach. Currently, QPack has implemented 6 VQAs, of which 4 are QAOA-based problems (Maximum Cut, Dominating Set, Maximal Independent Set and Traveling Salesperson problem) and 2 are VQE-based problems (Random Diagonal Hamiltonian and Ising Chain). See Appendix \ref{sec:algos} for their implementation. As proposed in an earlier version of the QPack benchmark~\cite{QPack_benchmark_paper}, quantum computer performance can be evaluated based on runtime, accuracy and scalability. This work proposes a quantitative evaluation of these performance categories, allowing for an unambiguous comparison between quantum computers. These categories each have their own sub-score, which is then later combined into an overall single-score metric. This provides an easy and quick comparison metric between quantum computers, but does not obscure the underlying characteristics of the quantum system too much, as sub-scores can hold valuable information about quantum computer performance and shed light on which areas can be improved.

This benchmark is built atop the cross-platform library LibKet~\cite{libket}, enabling the execution of the benchmark on multiple quantum computers with a single program. As QPack is created by an independent organization without its own commercially available quantum hardware, it is written in the spirit of full transparency and with the aim to enable a fair comparison of quantum computers without bias for a particular system or technology.
\section{Score approach} \label{sec:metrics}

This section covers the foundation of the QPack benchmark scores. The available quantum execution data collected by QPack is discussed and it is shown that differences in performance areas can be clearly observed. After covering the available quantum execution data, some criteria for benchmarking scores are defined to create scalable and intuitive scores.

\subsection{Quantum execution data}
During execution of a VQA, QPack takes multiple measurements  per execution cycle, resulting in a set of quantum execution data.  An example of such data can be found in the JSON snippet below, which shows the data for the first execution cycle of the QAOA MaxCut problem, with problem size 5, using the COBYLA optimizer~\cite{COBYLA}. Descriptions of all quantum execution data can be found in Table \ref{tab:metrics}.\\

\begin{verbbox}
{
 "Depth": 95.0,
 "Expectation Value": -5.9872612953186035,
 "Expectation Value Optimal": -6.0,
 "Job durations [ms]": [0.6613, 0.7702, 0.7714, ...],
 "Optimal params": [0.9055, 0.3867, ...],
 "Optimizer durations [ms]": [1288.003, 95.992, ...],
 "Optimizer iterations": 100,
 "Qubits": 5,
 "Total Classic duration [s]": 9.3162921,
 "Total Quantum duration [s]": 0.0727
}
\end{verbbox}
\resizebox{0.95\linewidth}{!}{\theverbbox}\\

\begin{figure}[h]
     \centering
     \begin{subfigure}[b]{0.49\linewidth}
        \centering
        \includegraphics[trim={0.0cm 14cm 13.0cm 1.1cm}, clip, width=\linewidth]{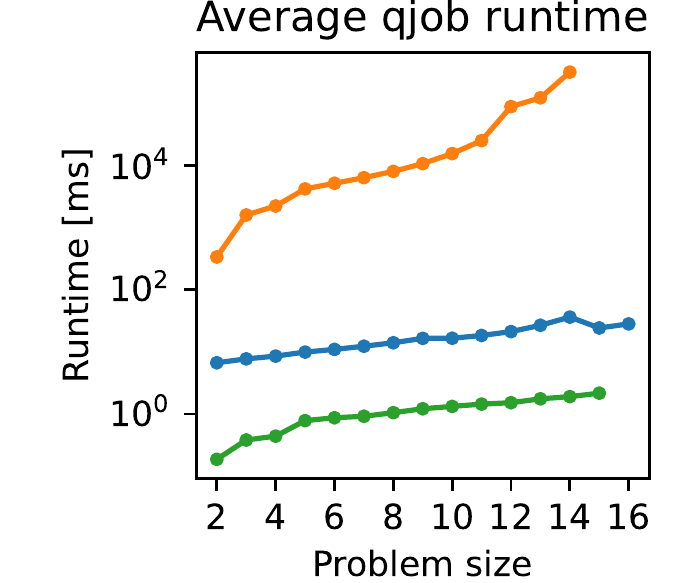}
        \caption{Quantum job time data}
        \label{fig:MCP_bench_qjob}
     \end{subfigure}
     \hfill
     \begin{subfigure}[b]{0.475\linewidth}
    \centering      
    \includegraphics[trim={0.0cm 7.6cm 13.35cm 7cm}, clip, width=\linewidth]{figures/MCP_benchmark.pdf}
    \caption{VQA error data}
    \label{fig:MCP_bench_qscore}
     \end{subfigure}
        \caption{Quantum execution data from the simulation of the MaxCut benchmark on the Rigetti QVM (orange), Qiskit Aer (blue) and QuEST (green) simulators. $(a)$: Average quantum job duration for a VQA run. $(b)$: Relative error between the quantum computer's expectation value and the baseline expectation value.}
        \label{fig:VQA_bench_example}
\end{figure}

This raw quantum execution data forms the basis of evaluating quantum computer performance. When plotted, this data can already reveal differences in execution characteristics  between quantum computer simulators, see the examples in Figure \ref{fig:MCP_bench_qjob} and \ref{fig:MCP_bench_qscore}. 

Here, a clear distinction can be made in different quantum computer qualities. For example, with a growing problem size, it can be seen that the QuEST simulator is the fastest and has the lowest error values among the tested quantum computers. These observations can now be processed, such that a quantitative comparison between quantum computers is possible. 

\subsection{Score criteria} \label{subsec:criteria}
Using the measurement data discussed in the previous section, performance scores can be distilled. For ease of comparison, a performance score should be a single "figure of merit", which allows for a clear distinction between different quantum computers, i.e., a quantum computer with a higher score is better than a quantum computer with a lower score. Nevertheless, scores should not become too abstract as to not properly reflect characteristics of a quantum computer. For instance, a quantum computer may be very fast but inaccurate, while another quantum computer has high accuracy but takes a long time to execute. Which one is then considered better? Other factors like scalability, maximum number of qubits, or serviceability also play a role in defining quantum computer performance.\\

\noindent
With this in mind, a number of criteria for benchmark scores are defined in the design of QPack:
 
\begin{itemize}
    \item Benchmark score reflects application-level performance of a quantum computer (simulators and hardware implementations)
    \item Benchmark score is a composite of measurement data of multiple quantum applications
    \item Benchmark score is a single number (but may be split up into sub-scores)
    \item Benchmark score is proportional to performance, i.e., a higher score means higher performance
    \item Benchmark score are scalable, i.e., score has no upper limit
    \item Benchmark score does not become too abstract from the data it is based on
    \item Sub-scores should be balanced, such that one sub-score does not become dominant in the overall score
\end{itemize}
\section{Benchmark Scores} \label{sec:score_def}

\begin{figure}[h]
    \centering
    \resizebox{0.8\linewidth}{!}{%
    \begin{tikzpicture}[grow=right,
  level 1/.style={level distance=2cm, sibling distance=2cm, anchor=west},
  level 2/.style={level distance=2cm, sibling distance=0.7cm, anchor=west}, 
  every node/.style = {shape=rectangle, rounded corners, draw, align=left, top color=white, bottom color=blue!20}]]
  \node {Overall Score}
    child { node {\textbf{Runtime}} 
        child{ node {Job durations} }
        child{ node {Depth} }
        child{ node {Shots} } }
    child { node {\textbf{Accuracy}}
        child{ node {Expectation value} }
        child{ node {Expectation value baseline} }}
    child { node {\textbf{Scalability}}
        child{ node {Job durations} }
        child{ node {Problem size} }}
    child { node {\textbf{Capacity}}
        child{ node {Expectation value} }
        child{ node {Baseline Expectation value} }
        child{ node {Threshold value} }
        child{ node {Problem size} }};
\end{tikzpicture}
}%
    \caption{Benchmark score decomposition. Each sub-score (bold) is connected to its relevant quantum execution data}
    \label{fig:score_decomp}
\end{figure}
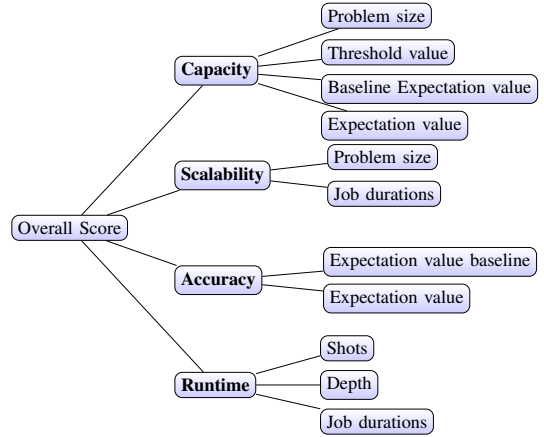

Using the aforementioned criteria, the actual benchmark scores can now be defined. Taking inspiration from BAPCo~\cite{bapco_2021, bapco_sysmark25}, an overall benchmark score can be decomposed into multiple sub-scores. These sub-scores and the connection between their quantum execution data (Table \ref{tab:metrics}) can be seen in Figure \ref{fig:score_decomp}. The overall score is divided into four sub-categories: \textit{runtime}, \textit{accuracy}, \textit{scalability} and \textit{capacity}. Runtime will evaluate the time the quantum computer needs to execute a given circuit. Accuracy reflects the ability of the VQA to find the optimal solution. Scalability evaluates the ability of the quantum computer to execute larger quantum circuit sizes. The capacity sub-score will reflect the number of qubits of a quantum computer for which the classical optimizer is able to find an optimal value below a predefined threshold.

For the benchmark sub-scores, a distinction will be made between pure scores $S^{\text{pure}}$ and mapped scores $S^{\text{mapped}}$. The pure scores are the values to which the measured data is transformed to a quantitative score metric. The mapped scores then take these pure scores and map them to be proportional to performance and be balanced against the other sub-scores.

\subsection{Runtime}
Perhaps the most straightforward metric to use is the time it takes to execute a quantum circuit. After all, quantum computing promises improved runtime of classical computers. For the runtime score, it is assumed that quantum computers can execute gates in parallel where possible. Then, to get a fair runtime score for different depths and shots, a score can be defined as the number of gates per second a quantum computer is able to execute. The number of gates $G$ for executing a single circuit depends on the circuit depth and the number of shots:

\begin{equation}
    G_{P,N} = D_{P,N} S_{P,N}
\end{equation}
\noindent
where $D_{P,N}$ and $S_{P,N}$ are the depth and number of shots for a given VQA problem $P$ and problem size $N$, respectively. Here, $D_{P,N}$ is the depth of the untranspiled circuit, that is, the hardware-agnostic implementation of QPack using the full set of software-visible gates. To ensure fair evaluation, this depth is the same for every quantum computer. This depth here is defined as the length of the critical path of the QPack untranspiled circuit. Transpiling the circuit in an efficient manner to fit to the qubit topology and its basis gate set is a job left for the quantum computer provider. Since this is an application-oriented benchmark, performance is evaluated on the quantum application that is executed, which will be reflected in the manner in which a circuit is transpiled by the quantum computer provider.

For a given VQA problem, the runtime score can thus be computed as the average gates per second over all problem sizes:

\begin{equation}
    S^{\text{pure}}_{\text{runtime}, P} =\frac{1}{N_{P,e} - N_{P,s}} \sum_{N=N_{P,s}}^{N_{P,e}} \frac{G_{P,N}}{T^{Q}_{P,N}}
\end{equation}
\noindent
where $T^{Q}_{P,N}$ is the average time of the quantum jobs for problem $P$ and problem size $N$. $ N_{P,s}$ and $N_{P,e}$  are the smallest and largest problem size for which the problem is evaluated.

\subsection{Accuracy}
Accuracy of the measured quantum state is an important aspect of quantum computers. Where fidelity is a common measure of single- or two- qubit gates, such low-level characteristics become more indistinguishable when the circuit size increases. The resulting output state deficiencies due to gate noise and qubit decoherence and relaxation can be compared to those of an ideal quantum simulator. For QPack, this entails comparing the performance of the classical optimizer on a perfect deterministic simulator to its performance on a noisy and nondeterministic quantum computer. In this case, the QuEST simulator is used as an ideal simulator with a deterministic output state. This way, the classical optimizer can perform in a noiseless case and find the lowest possible solution for a given problem. The reason that the theoretical achievable score is not used as a reference is the fact that not all VQAs optimally encode the solution. For QAOA, the number of circuit iterations $p$ limits the performance of QAOA efficiency. To not get circuits that grow too large, $p$ is static. In the case of VQE, the ansatz to find the ground state energy of the Hamiltonian does not always encode the complete solution space and thus limits accuracy of the VQE approximation.\\

\noindent
The accuracy score is defined as the average relative error between the expectation value of the ideal simulator (QuEST) and the quantum computer under test. The relative error is simply the absolute error divided by the ideal expectation value.

\begin{equation}
    S^{\text{pure}}_{\text{accuracy}, P} = \frac{1}{N_{P,e} - N_{P,s}} \sum_{N=N_{P,s}}^{N_{P,e}} \frac{E^{\text{ideal}}_{N,P}-E^{Q}_{N,P}}{E^{\text{ideal}}_{N,P}}
\end{equation}
\noindent
where $E^{\text{ideal}}_{N,P}$ and $E^{Q}_{N,P}$ are the ideal and quantum computer's minimal optimizer expectation values over all execution cycles, respectively.

\subsection{Scalability}
Another performance characteristic is the scalability of quantum circuits. Even though a quantum computer has a certain number of qubits to work with, its topology could make the evaluation of larger circuits more difficult as mapping and transpiling efficiently becomes more complex. Scalability looks at the runtime trend of a quantum computer with growing circuit size. 

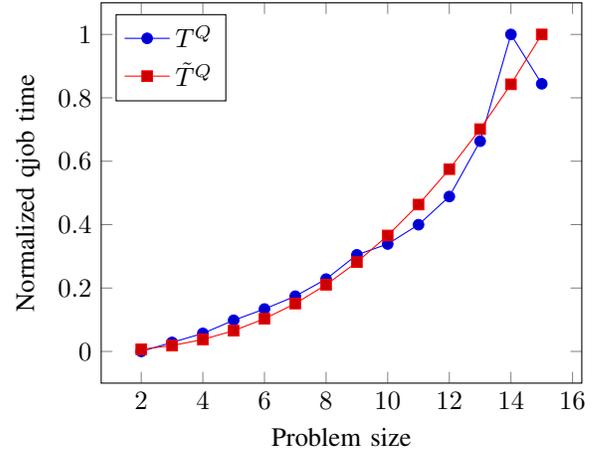
\begin{figure}[h]
    \centering
    \begin{tikzpicture}
    \begin{axis}[ylabel = Normalized qjob time,
                 xlabel = Problem size,
                 width = 0.9\linewidth,
                 height = 0.75\linewidth,
                legend pos = north west]
      \addplot coordinates {
        (2, 0)
        (3, 0.02847352)
        (4, 0.05692059)
        (5, 0.09827853)
        (6, 0.13380173)
        (7, 0.17420916)
        (8, 0.22810217)
        (9, 0.30464818 )
        (10, 0.33890984)
        (11, 0.399752)
        (12, 0.48869774 )
        (13, 0.66320361)
        (14, 1.)
        (15, 0.84423322)
    };
    \addplot coordinates {
        (2, 2^2.48059946 / 15^2.48059946)
        (3, 3^2.48059946 / 15^2.48059946)
        (4, 4^2.48059946 / 15^2.48059946)
        (5, 5^2.48059946 / 15^2.48059946)
        (6, 6^2.48059946 / 15^2.48059946)
        (7, 7^2.48059946 / 15^2.48059946)
        (8, 8^2.48059946 / 15^2.48059946)
        (9, 9^2.48059946 / 15^2.48059946)
        (10, 10^2.48059946 / 15^2.48059946)
        (11, 11^2.48059946 / 15^2.48059946)
        (12, 12^2.48059946 / 15^2.48059946)
        (13, 13^2.48059946 / 15^2.48059946)
        (14, 14^2.48059946 / 15^2.48059946)
        (15, 15^2.48059946 / 15^2.48059946)
    };

    \legend{
    $T^{Q}$,
    $\tilde{T}^{Q}$
    }
    \end{axis}
\end{tikzpicture}
    \caption{Normalized input data and fitted curve for MaxCut problem on Qiskit Aer simulator}
    \label{fig:scale_curves}
\end{figure}

To quantify the scalability, the exponential growth of the average quantum job time against the problem size will be evaluated. This is done by fitting the quantum job time $T^{Q}$ to function:

\begin{equation}
    \tilde{T}^{Q}(N) = N^a
\end{equation}
\noindent
where $N$ is the problem size of the current problem. To fit this curve, the value of $a$ needs to be determined. This is done by normalizing the input data and optimize the value of $a$ with a least-squares cost function. This way, the best fit for $a$ can be found for the input data, as shown in Figure \ref{fig:scale_curves}.

The value of $a$ can now be used as a quantification of the scalability score, see Equation \ref{eq:scalabilty_pure}. For example, if a quantum job time scales linearly with the problem size, $a=1$ is expected. In the case that $a > 1$, the qjob time grows faster than the problem size and vice versa for $a < 1$.

\begin{equation} \label{eq:scalabilty_pure}
    S^{\text{pure}}_{\text{scalability}, P} = a_P
\end{equation}

\subsection{Capacity}
The final score metric is the capacity of the quantum computer. This is the number of qubits that a quantum computer is able to run within a margin of the desired output accuracy. This is a similar approach as the Atos Q-score~\cite{Atos_benchmark}, but generalized over multiple quantum applications and using the QuEST simulator with the classical optimizer as a baseline. The capacity score is then the highest number of qubits $Q_N$ corresponding to problem size $N$ for which the quantum computer can achieve a relative error within a set threshold accuracy $A^*$. The score is thus defined as:

\begin{equation}
    S^{\text{pure}}_{\text{capacity}, P} = \text{max}\{Q_N \mathrm{\ where\ }  \frac{E^{\text{ideal}}_{N,P}-E^{Q}_{N,P}}{E^{\text{ideal}}_{N,P}} \leq A^*\}
\end{equation}

The main concern for this benchmark score is the value of $A^*$, which is chosen arbitrarily. Currently, this is implemented as 20\% relative accuracy. 

\subsection{Sub-score mapping, balancing \& combining} 

With all pure sub-score metrics defined for each problem set, they need to be mapped and balanced to create scores that operate in a similar range to one another. Mapping transforms the pure score, such that a better performance in a score category leads to an increased benchmark score. This is then scaled up or down in order for the sub-score to be in a similar range relative to the other sub-scores, which we refer to as balancing the sub-scores.

For the runtime sub-score, the decimal logarithm function is used to scale down the pure sub-score

\begin{equation}
    S^{\text{mapped}}_{\text{runtime}, P} = log_{10}(S^{\text{pure}}_{\text{runtime}, P})
\end{equation}
\noindent
which is able to map the whole range of the pure sub-score, because $S^{\text{pure}}_{\text{runtime}, P} > 0$.\\

Then, the accuracy score can be scaled to a similar range as the runtime mapped score. However, taking a log function is not applicable, since $S^{\text{pure}}_{\text{accuracy}, P} \in \mathbb{R}$, e.g. the possibility that a simulator performs better than the baseline is allowed. Since a larger sub-score should represent a better solution, a lower $S^{\text{pure}}_{\text{accuracy}, P}$ should output a higher score. This can be achieved with the mapping function

\begin{equation}
    f_{\text{map}}(x) = \frac{\pi}{2} - \text{arctan}(x)
\end{equation}
\noindent
which maps the value $x$ into the range $[0, \pi ]$ as $x$ decreases. Using this mapping function, we can map and balance the accuracy sub-score to

\begin{equation}
    S^{\text{mapped}}_{\text{accuracy}, P} = c_0 f_{\text{map}}(c_1 S^{\text{pure}}_{\text{accuracy}, P})
\end{equation}
\noindent
where $c_0$ scales the mapping function and $c_1$ adjusts the sensitivity of the mapping function. By trial and error, $c_0 = \frac{30}{\pi}$ and $c_1 = 50$ were found to be adequate.\\

The same mapping function can be used for the scalability sub-score, because again, a lower value for $S^{\text{pure}}_{\text{scalabilty}, P}$ should result in a higher score. Although $S^{\text{pure}}_{\text{scalabilty}, P} \in \mathbb{R}$, values of $S^{\text{pure}}_{\text{scalabilty}, P}$ are mostly expected to be larger then one. To account for this, the mapping function is shifted by one:

\begin{equation}
    S^{\text{mapped}}_{\text{scalability}, P} = c_2 f_{\text{map}}(c_3( S^{\text{pure}}_{\text{scalabilty}, P}-1))
\end{equation}
\noindent
where $c_2$ scales the mapping function and $c_3$ adjusts the sensitivity of the mapping function. By trial and error, $c_2 = \frac{30}{\pi}$ and $c_3 = 0.75$ were found to be adequate.\\

The capacity score is found to already be in a similar range to the other performance metrics and reflects the performance of a quantum computer adequately. The capacity score is then kept as is:

\begin{equation}
    S^{\text{mapped}}_{\text{capacity}, P} = S^{\text{pure}}_{\text{capacity}, P}
\end{equation}

Now that a mapped and balanced sub-score has been defined for each problem, sub-scores for each problem need to be combined. For a quantum computer, each sub-score $S$ for each performance category is computed as the arithmetic mean over all problems:

\begin{equation} \label{eq:arith_mean_P}
    S = \frac{1}{n_P} \sum_P S^{\text{mapped}}_P
\end{equation}
\noindent
where $n_P$ is the number of problems that the quantum computer is evaluated on.

\begin{figure}[h]
    \centering
    \includegraphics[width = \linewidth]{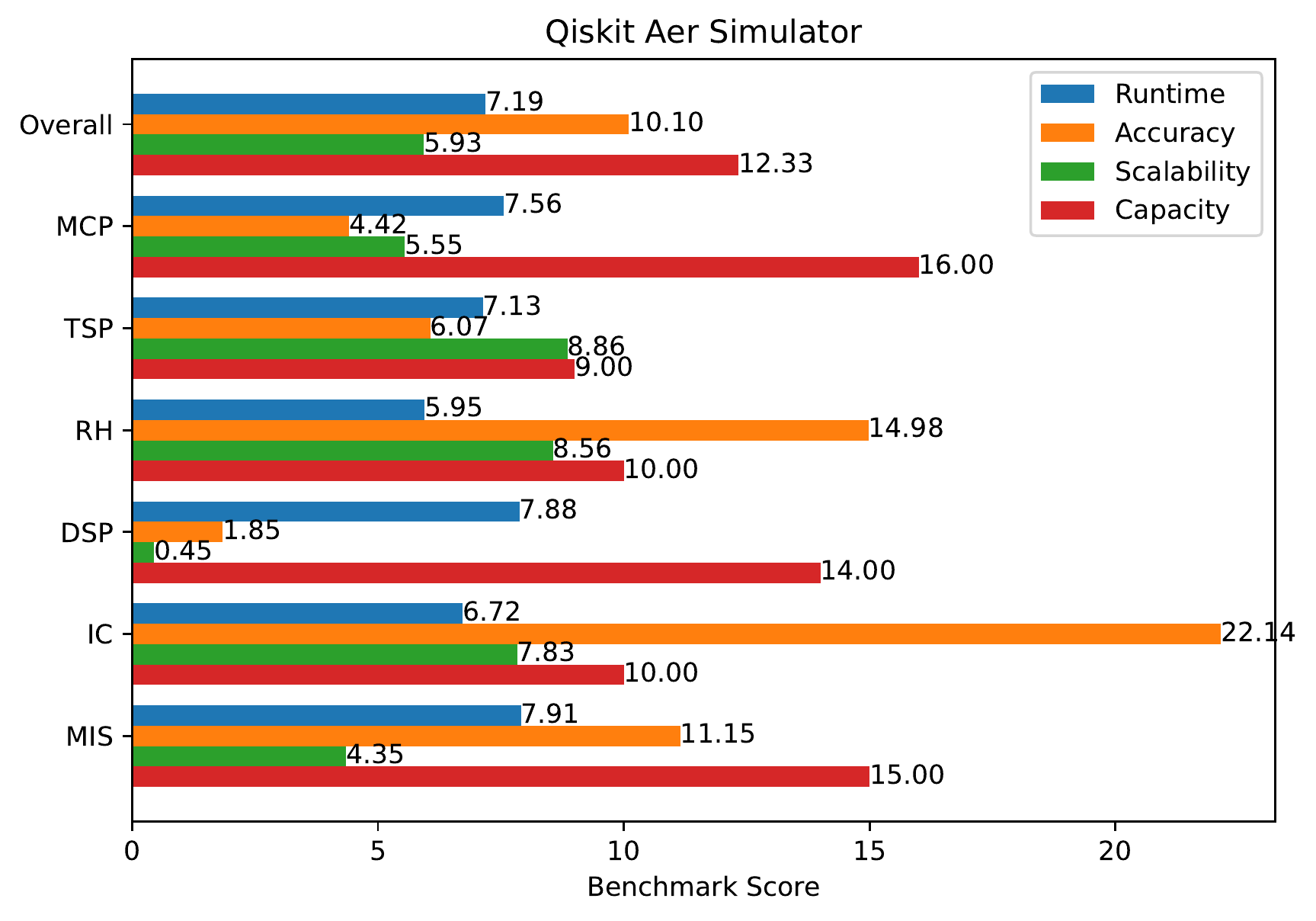}
    \caption{QPack result for the Qiskit Aer local simulator}
    \label{fig:Qiskit_Aer_result}
\end{figure}

For every quantum computer, sub-scores for each problem in the problem set are computed, after which the combined sub-scores are derived using Equation \ref{eq:arith_mean_P}. An example of scores obtained can be seen in Figure \ref{fig:Qiskit_Aer_result}.

\subsection{Overall score}

\begin{figure}[h]
    \centering
    \includegraphics[width = 0.9\linewidth]{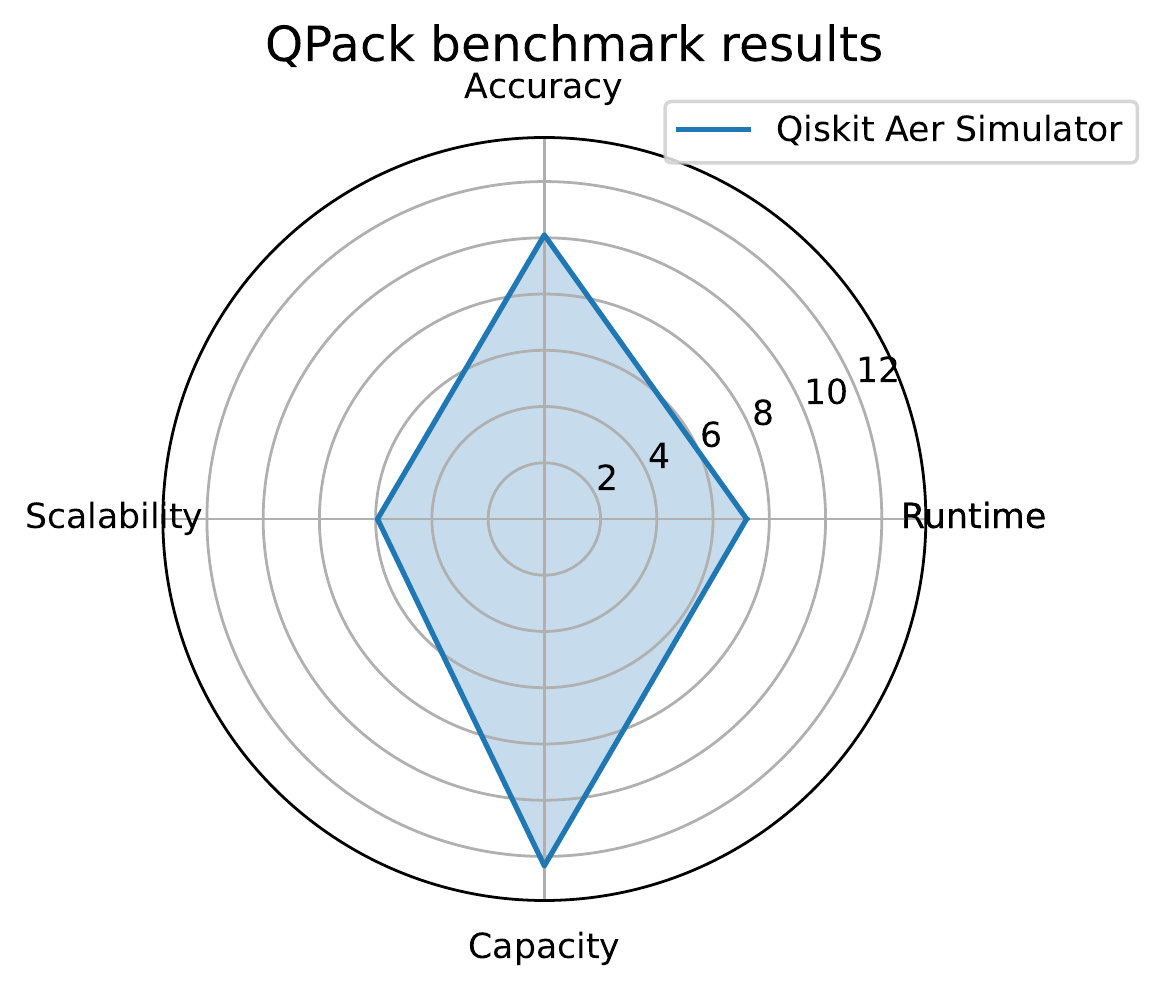}
    \caption{Benchmark result radar plot for the Qiskit Aer simulator}
    \label{fig:sim_results}
\end{figure}

Computing the sub-scores for a single quantum computer and displaying them like Figure \ref{fig:Qiskit_Aer_result} gives a good insight of the performance of a single computer, but makes comparison between multiple quantum computers cumbersome. A better way of visualizing performance differences between quantum computers can be done with radar plots. Such a plot is shown in Figure \ref{fig:sim_results}.

Using this visualization, we can provide a single parameter to compare different compute platforms by taking the area of the four-sided region in Figure \ref{fig:sim_results}. This defines the overall score as:

\begin{equation}
    S = \frac{1}{2}(S_{\text{runtime}} + S_{\text{scalability}})(S_{\text{accuracy}} + S_{\text{capacity}})
\end{equation}

This overall score fulfills the performance proportionality requirement and provides a simple way to compare between compute platforms. The advantages of this overall score are its simplicity and general applicability.

\section{Results} \label{sec:results}
The proposed benchmark scores can now be applied to various quantum simulators and actual quantum hardware. This section shows the results for simulators which can be installed on a local machine, cloud-accessible simulators and cloud-accessible quantum computers. Only the final sub-scores and overall scores of the units under test are shown. For all quantum execution data sets and benchmark scores per problem, see Appendix \ref{app:datasets} and \ref{app:bench_result}, respectively.

\subsection{Local simulators}
\begin{figure}[h]
    \centering
    \includegraphics[width = 0.9\linewidth]{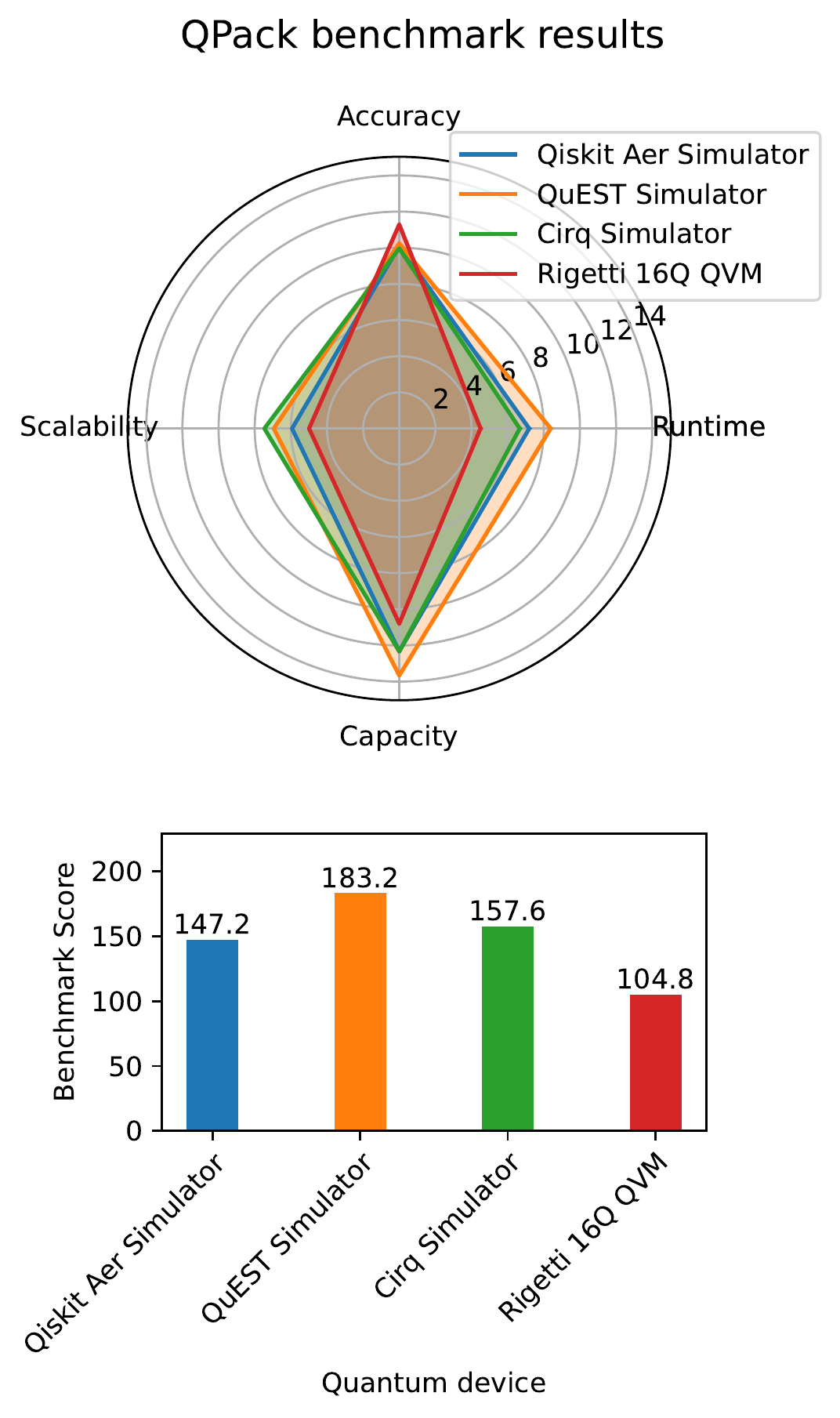}
    \caption{Score comparison for evaluated local simulators}
    \label{fig:loc_sim_result}
\end{figure}

The local quantum computer simulators tested in this paper are the Qiskit Aer~\cite{qiskit_aer}, Cirq~\cite{cirq_simulator}, Rigetti QVM~\cite{rigetti_qvm} and QuEST~\cite{quest} simulators without noise models.  Data was collected using a Windows 10 desktop computer, utilizing an AMD Ryzen 5 3600 6-core CPU~\cite{ryzen5} with 16 GB RAM in an Ubuntu Windows Subsystem for Linux environment. Their benchmark scores are shown in Figure \ref{fig:loc_sim_result}, where sub-scores are plotted in the top radar plot and the overall scores are shown in the bottom bar graph.

From this radar plot and the overall scores below them, it can be seen that the QuEST simulator achieves the highest overall score and gets the highest sub-scores in runtime and capacity. The Cirq simulator performs best in terms of scalability. All simulators seem to perform similarly in terms of accuracy, with the Rigetti QVM simulator having the best accuracy performance. Notice that although the Rigetti QVM simulator scores the highest in accuracy, it does perform the worst in capacity. This can be explained by the fact that this simulator was not able to complete larger problem sizes as runtime increased rapidly as the problem size grew, which is also reflected in the fact the the Rigetti QVM has the worst runtime and scalability scores.

The Cirq and Qiskit Aer simulators have a close overall score. They have comparable scores in accuracy and capacity, but differ in runtime and scalability. The Cirq simulator has a lower runtime score than the Aer simulator, but it has a better scalability score. This may be a reason to favor Cirq over Qiskit if scalability is of more concern than actual runtime. 

\subsection{Remote simulators}
\begin{figure}[h]
    \centering
    \includegraphics[width = 0.9\linewidth]{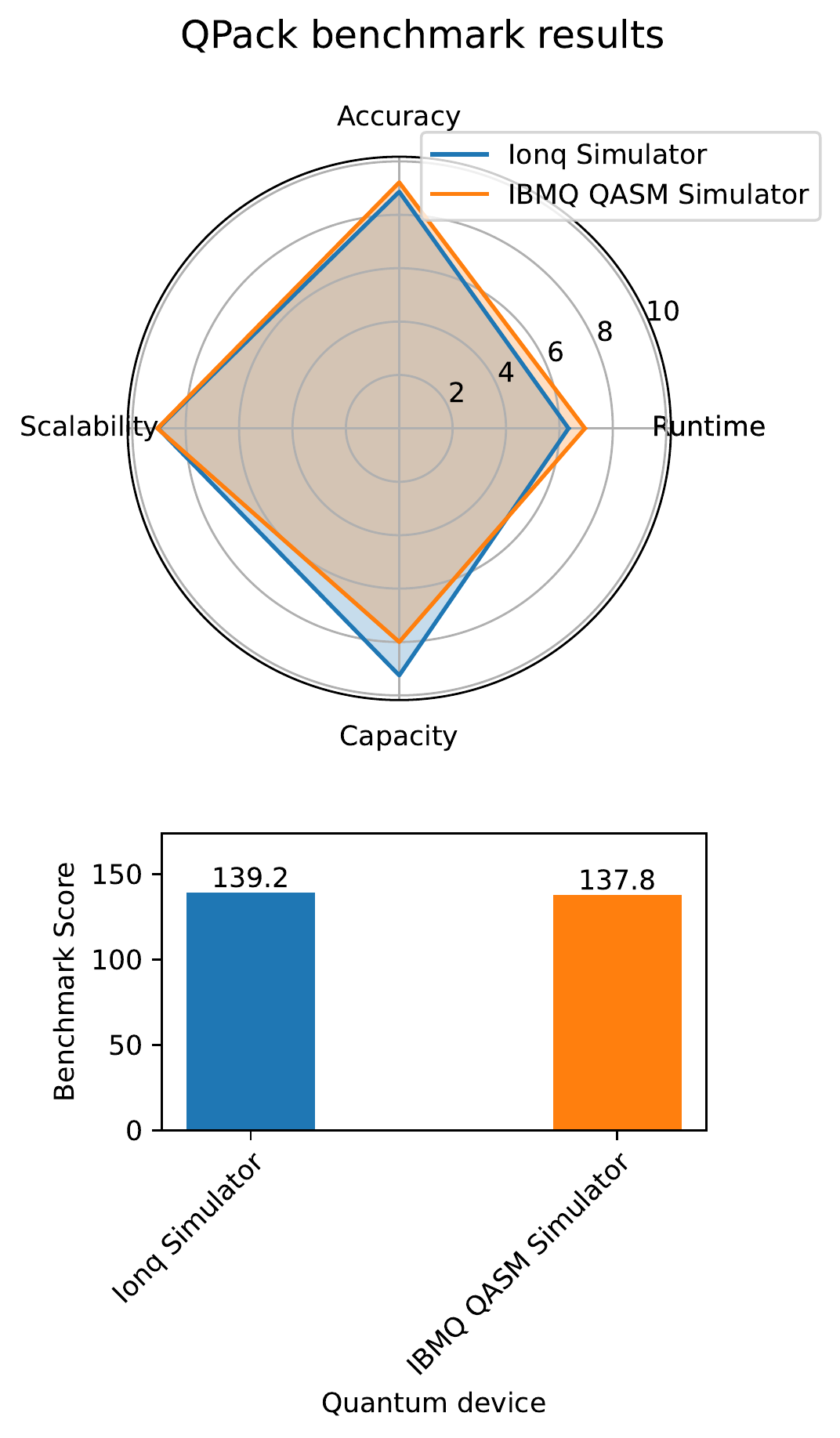}
    \caption{Score comparison for evaluated remote simulators}
    \label{fig:rem_sim_result}
\end{figure}

If local quantum simulation is not preferred, many providers allow cloud-access to their simulators, such as IBMQ, IonQ and Rigetti. In this work, a comparison is made between the IBMQ QASM simulator~\cite{ibmq_qasm_simulator} and the IonQ simulator~\cite{ionq_simulator}, shown in Figure \ref{fig:rem_sim_result}.

Here it is shown that both simulators have very similar performance. For overall quantum performance, QPack ranks the IonQ simulator  as the more superior simulator of the two, although differences are minor. The IBMQ QASM simulator is faster and more scalable then the IonQ simulator. However, it should be noted that the serviceability of remote quantum computers could also be an important factor to choose a quantum provider. For general public access, the IonQ usually has a faster response time than the IBMQ QASM simulator, as a result of smaller queues. IonQ achieves a higher capacity score, but this is mainly due to the fact that IonQ has completed a larger set of problem sizes than the IBMQ QASM simulator.

\subsection{Remote hardware}
\begin{figure}[h]
    \centering
    \includegraphics[width = 0.9\linewidth]{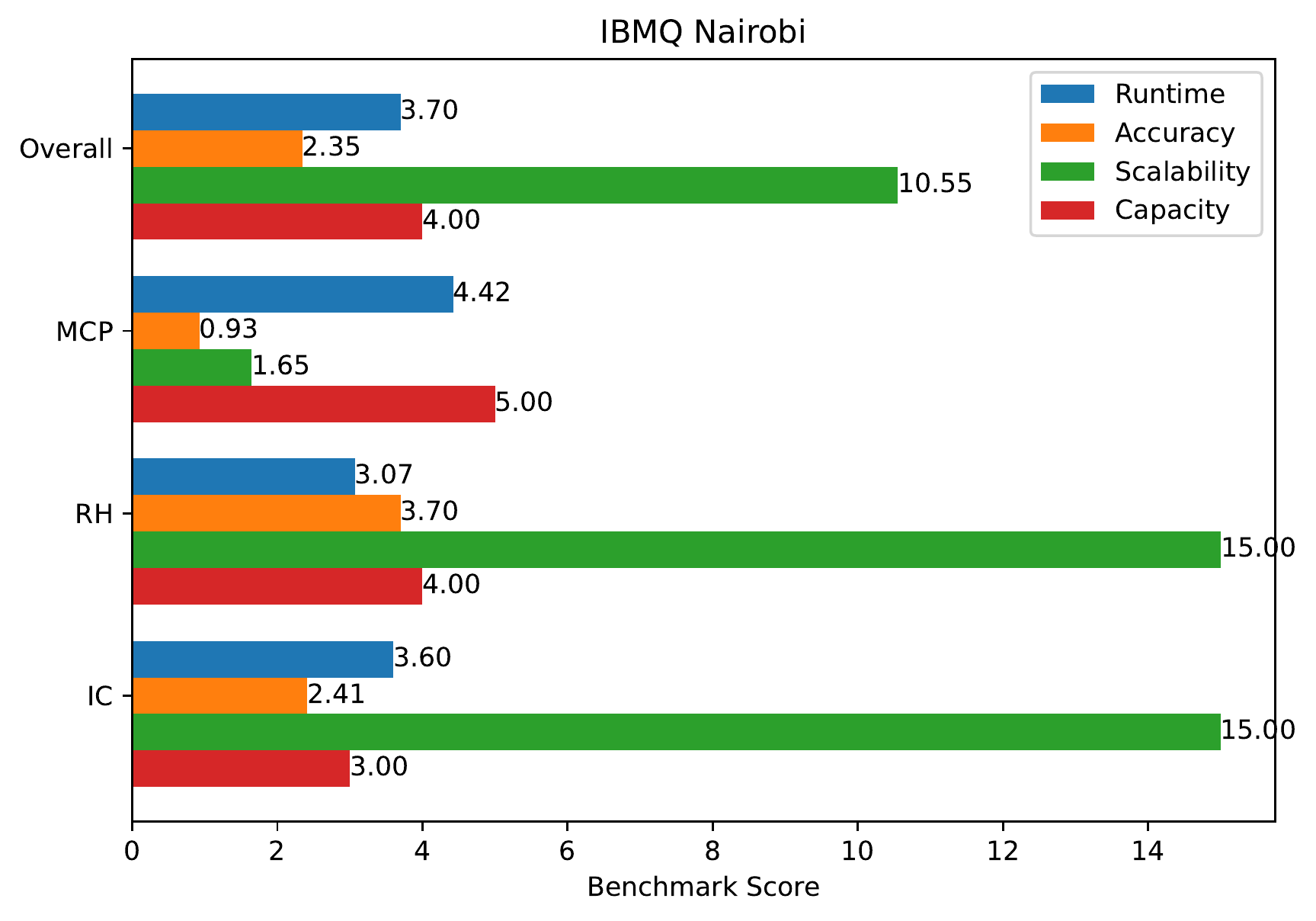}
    \caption{QPack benchmark result for the IBMQ Nairobi 7-qubit quantum processor}
    \label{fig:rem_qpu_result}
\end{figure}

Currently the only hardware tested in this paper is the 7-qubit IBMQ Nairobi quantum processor~\cite{ibm_aviary}. At this moment, only a limited set of data has been acquired, making a complete evaluation of the Nairobi processor not yet possible. Results collected thus far can be found in Appendix \ref{app:datasets}, which results in the following scores for the QAOA MaxCut problem, and the Random Hamiltonian and Ising chain VQE problems, as shown in Figure \ref{fig:rem_qpu_result}. 

Notice that due to the limited set, it seems like the Nairobi scores very well in scalability, but due to the limited problem size set to fit the pure scalability score, this evaluation is currently not a good representation of the actual runtime trend. This stresses the need for a further collection of quantum execution data and this will be presented in a future version of this paper.

\section{Discussion} \label{sec:discussion}

The presented results indicate that the QPack benchmark is able to give an easy and insightful comparison between quantum computers. However, it should be noted that these scores use some arbitrary values. One of these, for example, is the threshold value $A^*$, which could be chosen to be more or less tolerant to a given output state. In this implementation, the value of $A^*$ was chosen such that most local simulators without noise models would achieve a maximum capacity score for most problem sets to compensate for their nondeterministic behavior. This sets the bar at such a level that a real quantum computer can reach a similar output performance as their simulated noiseless counterparts.

The balancing parameters $c_0$, $c_1$, $c_2$, $c_3$ are also subject to discussion, as changing their values could change the accuracy and scalability sub-scores drastically. They were chosen such that different score categories fall within a similar range, making comparison and visualization more practical. Scores are thus not absolute values, but can give a relative difference between two quantum computers, as long as parameter choices are consistent. Related to this is the combination of the sub-scores to form the overall score. This is a practicality to support the ability of QPack to compare quantum hardware with a single performance metric. Again, consistency is key for a fair comparison between quantum computers.

Another impact on the benchmark scores is the set of collected problem data. For example, when comparing a 5- and 7-qubit quantum computer, a larger problem set can be run on the 7-qubit quantum computer. For a fair comparison, both quantum computers should be evaluated on the same problems for the same problem sizes. However, this could punish the 7-qubit device as it is theoretically capable of achieving a higher capacity score, but as a result its overall accuracy score possibly decreases. This may be a problem especially for NISQ-era quantum computers, as differences between running problem sizes that only differ by a small amount can have a notable impact on the benchmark performance. As quantum computers become larger, this problem will likely be less prominent.



It should be noted that the the runtime sub-score is now purely based on the execution time of the quantum circuit and does not include other service overhead like transpiling, scheduling and optimization. Although this gives an insight into the performance of the actual quantum hardware, it does not reflect other runtime factors of the whole system. Currently, overhead is a major component in overall quantum job runtime and is therefore also contributing to the runtime performance of a quantum computer. Since quantum providers generally only return the quantum execution time in their response, it was chosen to only include this runtime value in the current version of QPack. However, since the impact of this overhead has such an significant influence on quantum runtime, the next version of QPack should be able to evaluated these runtime values as well.

\section{Conclusion} \label{sec:conclusion}
In this paper, the quantum benchmarking scores of QPack were presented. After an overview of the current state of quantum benchmarking, opportunities were made clear for a more holistic and versatile benchmarking approach, using VQAs as an application-level benchmark to create a variety of scalable benchmarking circuits. Running these benchmarking circuits on different quantum computers using the cross-platform library LibKet allows for a collection of quantum execution data to be used as a quantum performance indicator. 

The transformation of the raw quantum execution data to the QPack scores provides four performance sub-scores based on runtime, accuracy, scalability and capacity. Runtime evaluates how fast a quantum computer can execute a quantum circuit on average. Accuracy then checks how well the output of a quantum computer performs compared to an ideal quantum computer simulator. Scalability tells us something about how well the quantum computer is able to handle increasing circuit sizes. Finally, the capacity score indicates what the usable number of qubits on a device actually is, no matter how many qubits a quantum computer can provide.

These quantitative performance metrics provide a holistic and varied method of benchmarking quantum computers. Benchmark scores allow for quick and easy comparisons between quantum computers and show in what areas performance gain is possible. 

A proof-of-concept has been provided by running QPack on a set of local and remote simulators, as well as a single quantum processor. Local simulators from Cirq, Qiskit, Rigetti and QuEST have been compared to one another, where it was shown that the QuEST simulator has the best overall performance. The remote simulators of IBMQ and IonQ have been compared and have shown to have a similar performance in the QPack benchmark. A small set of quantum execution data has also been collected for the IBMQ Nairobi quantum processor to verify that the QPack benchmark is also applicable for quantum hardware.


As shown in Section \ref{sec:metrics}, only a few of the presented quantum execution data types are used to compute the QPack scores. This indicates that there are possibilities to create more sub-scores. For example, a comparison between classical and quantum runtimes could be made or the serviceability of a remote quantum computer can be quantified in a new sub-score. 

Further improvements to QPack could be made by implementing more problem sets of different VQAs or non-VQAs. Algorithms such as HHL~\cite{HHL}, Hydrogen simulation~\cite{H2_Sim} or Shor's algorithm~\cite{Shor_alg} could offer viable NISQ-era quantum applications to use in the further development of QPack, allowing for an even more varied set of quantum circuits to evaluate quantum performance.

With the performance scores tested on a variety of local and remote quantum simulators, the QPack benchmark aims to evaluate more actual quantum computers next to the IBMQ Nairobi. From the IBMQ family, the Jakarta, Lagos, Quito, Manila and Perth quantum processors are planned to be evaluated, as well as quantum computers from Rigetti, Honeywell and Quantum Inspire. This should give an insightful comparison between quantum computers of different vendors and could show what the current state of quantum technology is at this time. 

The link to the latest version of QPack and the collected quantum execution data can be found using the following GitLab Link: \url{https://gitlab.com/libket/qpack}

\bibliography{bibliography.bib}

\begin{appendices}
\section{VQE \& QAOA general} \label{app:vqa_math}

This section describes the variational quantum algorithms that are implemented in the QPack benchmark. A review and main purpose of each algorithm is briefly stated.

Variational Quantum Algorithms (VQAs) have been proposed as a leading strategy to work within the constraints of the NISQ quantum devices. VQAs use a classical optimizer to train a parameterized quantum circuit. According to~\cite{VQA}, these algorithms will pave the way for all applications that quantum computers were envisioned for. The main idea is that these parameterized circuits are small and thus still give meaningful results which can be processed by a classic computer.\\

\subsection{VQE}
The first VQA that was proposed was the Variational Quantum Eigensolver (VQE) in April 2013~\cite{VQE}. This algorithm can be used to find the eigenvectors of a Hamiltonian. The algorithm consists of two main parts. The first part is the quantum device that computes the (partial) expected values of the Hamiltonian $\mathcal{H}$ for any input state $\ket{\psi(\vec{\theta})}$. The second part consists of a classical computer which evaluates the expectation value. It can then alter the input state $\ket{\psi(\vec{\theta})}$, by tweaking the ansatz parameters $\vec{\theta}$ such that it minimizes
\begin{equation*}
    \frac{\bra{\psi(\vec{\theta})}\mathcal{H}\ket{\psi(\vec{\theta})}}{\braket{\psi(\vec{\theta})}}
\end{equation*}

By varying $\ket{\psi(\vec{\theta})}$ one may find unknown eigenstates. The parameters that minimize these eigenvalues are simply stored to generate these eigenstates when needed.

Another approach to use the VQA method to find eigenvalues was presented in 2020; the Contracted Quantum Eigensolver~\cite{CQE}. The main problem tackled was that VQE suffers from optimization over a surface, which is nonideal for high-dimensional optimization. This increases the computational load significantly as the system grows. The main idea is the same as the VQE, but the subspace is restricted, allowing for a more efficient optimization.

\subsection{QAOA}
\subsubsection{Original algorithm}
This algorithm is used by Atos~\cite{Atos_benchmark} and QPack~\cite{QPack_benchmark_paper}. The Quantum Approximate Optimization Algorithm (QAOA) was introduced by Fahri. et al in 2014~\cite{QAOA_fahri}. The QAOA-algorithm consists of two main parts: the cost-unitary and the mixer unitary. The cost function is defined as:

\begin{equation}
    C(z) = \sum_{\alpha = 1}^m C_{\alpha}(z)
\end{equation}
\noindent
where $z = z_1 z_2 ... z_n$ is the bit string to evaluate, $m$ the number of clauses and $C_\alpha (z) = 1$ if $z$ satisfies clause $\alpha$ and 0 otherwise. This cost function is then used as an operator in an unitary operator:

\begin{equation}
    U(C, \gamma) = e^{-i\gamma C} = \prod_{\alpha = 1}^m e^{-i\gamma C_\alpha}
\end{equation}
\noindent
which depends on angel $\gamma \in [0, 2\pi]$. The mixer unitary is simply the sum of single qubit $\sigma^x$ operators. The unitary operator can thus be defined as:

\begin{equation}
    U(B, \beta) = e^{-i \beta B} = \prod_{j=1}^n e^{-i\beta \sigma^x_j} 
\end{equation}
\noindent
with $\beta \in [0,\pi]$ and $n$ the number of qubits. These unitaries are then applied to an initial state in superposition: $\ket{s} = \frac{1}{\sqrt{2^n}}\sum_z \ket{z} = \ket{+}_1\ket{+}_2 ... \ket{+}_n$. The cost and mixer unitary can then be applied $p$ times to the initial state, giving the QAOA function the following quantum state

\begin{equation}
    \ket{\vec{\gamma}, \vec{\beta}} = U(B, \beta_p)U(C, \gamma_p) ... U(B, \beta_1)U(C, \gamma_1)\ket{s}
\end{equation}
\noindent
where $\vec{\gamma} = (\gamma_1, \gamma_2, ..., \gamma_p)$ and $\vec{\beta} = (\beta_1, \beta_2, ..., \beta_p)$. This means that the QAOA algorithm will always have $2p$ parameters. Now the quantum algorithm has been changed into an classical optimization function where the cost function can be maximized by finding the optimal parameters $\vec{\gamma}$ and $\vec{\beta}$. The expectation value of $C$ is then simply

\begin{equation}
    F_p = (\vec{\gamma}, \vec{\beta}) = \bra{\vec{\gamma}, \vec{\beta}}C\ket{\vec{\gamma}, \vec{\beta}}
\end{equation}
\noindent
and the maximum value
\begin{equation}
    M_p = \max_{\vec{\gamma}, \vec{\beta}} F_p(\vec{\gamma}, \vec{\beta})
\end{equation}

Note that 
\begin{equation} \label{eq:meerpmeerbeter}
    M_p \leq M_{p-1}
\end{equation}
\noindent
and thus increasing $p$ will generally result in a higher maximum result. However, for NISQ era quantum hardware, only small values for $p$ are generally used.

\subsubsection{Extended algorithm}
Five years after the introduction of QAOA, an extension of the algorithm was proposed, named the \textit{Quantum Alternating Operator Ansatz} to keep the abbreviation the same~\cite{QAOA_ansatz}. This extension allows for the algorithm to have a more varied set of states than the original. The paper takes a more general approach to optimization. 

An \textit{optimization problem} is a pair $(F, f)$, where $F$ is the domain and $f:F \rightarrow \mathbb{R}$ is the objective function. In quantum terms, this corresponds to $\mathcal{F}$ being the Hilbert space of dimension $|F|$, with standard basis $\{\ket{x}: x \in F \}$. The QAOA function in general consists of two main parts:

\begin{itemize}
    \item Phase-separation operators $U_P(\gamma)$, depending on the objective function $f$
    \item Mixing operators $U_M(\beta)$, depending on the domain and its structure
\end{itemize}

The QAOA algorithm then exists of three parts: The phase-separation operators, mixing operators and a starting state. This extension allows for more than just Hamiltonian based cost functions and mixers and can be useful to map more complex problems to the QAOA algorithm.

\section{Benchmark applications} \label{sec:algos}
Variational Quantum Algorithms (VQAs) have been proposed as a leading strategy to work within the constraints of the NISQ quantum devices. VQAs use a classical optimizer to optimize the output of a parameterized quantum circuit. According to~\cite{VQA}, these algorithms will pave the way for all applications that quantum computers are envisioned for. The main idea is that these parameterized circuits are small (few noise sources) and thus still give meaningful results which can be processed by a classic computer. This makes them a suitable set of quantum applications to be run on near-term devices and simulators. For QPack, two popular VQAs are used for different problem sets: The Quantum Approximate Optimization Algorithm (QAOA) and the Variational Quantum Eigensolver (VQE).

\subsection{QAOA}

This chapter will focus on the implementation of four networking problems to solve using QAOA implemented in the LibKet library. All problems will briefly be described, after which a mapping to QAOA will be presented. All problems are graph-based NP-hard problems~\cite{complexity_summary, VQAISNPHARD}.

\begin{figure}[h]
    \centering
    \begin{tikzpicture}
        \node[shape=circle,draw=black,scale=0.75] (A) at (0.6,0) {0};
        \node[shape=circle,draw=black,scale=0.75] (B) at (0.1,1.5) {1};
        \node[shape=circle,draw=black,scale=0.75] (C) at (1.5,2.5) {2};
        \node[shape=circle,draw=black,scale=0.75] (D) at (2.9,1.5) {3};
        \node[shape=circle,draw=black,scale=0.75] (E) at (2.4,0) {4};
    
        \path [-](A) edge node {} (B);
        \path [-](A) edge node {} (C);
        \path [-](B) edge node {} (C);
        \path [-](B) edge node {} (D);
        \path [-](C) edge node {} (D);
        \path [-](C) edge node {} (E);
        \path [-](D) edge node {} (E);
        \path [-](D) edge node {} (A);
        \path [-](E) edge node {} (A);
        \path [-](E) edge node {} (B);

    \end{tikzpicture}
    \caption{A 4-regular graph with 5 nodes}
    \label{fig:regular_graph}
\end{figure}
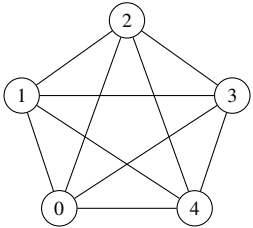

The network topology for a graph problem will be a 4-regular graph, i.e a connected graph where all nodes have degree 4. To increase the problem size, the number of nodes in the regular graph can simply be increased. This will be useful later on as the QPack benchmark will measure the performance of quantum systems by increasing the problem size. An illustration of such a graph is given in Figure \ref{fig:regular_graph}.

\subsubsection{MaxCut Problem}
The MaxCut problem is often used as a problem when QAOA is introduced. Although the problem is classified as NP-Complete~\cite{complexity_summary}, the problem is simple and straightforward to explain. The MaxCut problem was also used as an example when Fahri. et al. introduced QAOA~\cite{QAOA_fahri} and was proven to work well for low $p$ iterations. \\

The MaxCut problem is defined as follows. Given a graph $G=(V,E)$, with $V$ nodes (vertices) and $E$ links (edges), find the subset $S$ of $V$, such that the links between $S$ and $V \backslash S$ is maximized. In other words, how can one divide the nodes of the graph in two distinct sets, such that if one would cut the links between those sets, the amount of cut links is the largest possible. 

\begin{figure}[h]
    \centering
    \tikzset{
    phase label/.append style={above right,xshift=-0.1cm,yshift=0.1cm}
    }
    \resizebox{\linewidth}{!}{%
        \begin{quantikz}[column sep=0.2cm, row sep=0.1cm]
    \lstick{$q_0$} & \gate{H} &\ctrl{1}    &\ctrl{2}    &\qw         &\qw         &\qw         &\qw         &\qw         &\ctrl{3}    &\ctrl{4}    &\qw         & \gate{R_x} & \meter{} \\
    \lstick{$q_1$} & \gate{H}             &\rzz{\gamma}&\qw         &\ctrl{1}    &\ctrl{2}    &\qw         &\qw         &\qw         &\qw         &\qw         &\ctrl{3}    & \gate{R_x} & \meter{} \\
    \lstick{$q_2$} & \gate{H}             &\qw         &\rzz{\gamma}&\rzz{\gamma}&\qw         &\ctrl{1}    &\ctrl{2}    &\qw         &\qw         &\qw         &\qw         & \gate{R_x} & \meter{} \\
    \lstick{$q_3$} & \gate{H}             &\qw         &\qw         &\qw         &\rzz{\gamma}&\rzz{\gamma}&\qw         &\ctrl{1}    &\rzz{\gamma}&\qw         &\qw         & \gate{R_x} & \meter{} \\
    \lstick{$q_4$} & \gate{H}             &\qw         &\qw         &\qw         &\qw         &\qw         &\rzz{\gamma}&\rzz{\gamma}&\qw         &\rzz{\gamma}&\rzz{\gamma}& \gate{R_x} & \meter{} 
    \end{quantikz}%
    }
    \caption{Circuit for the 5 node regular 4-graph example. The circuit shows the QAOA implementation for $p=1$. The $ZZ$-gates describe a $ZZ$-rotation (see Appendix \ref{app:gate_double_paulis}) by angle $\gamma$ and the $R_x$-gates depict an X-rotation by angle $\beta$.}
    \label{fig:MCP_circuit}
\end{figure}
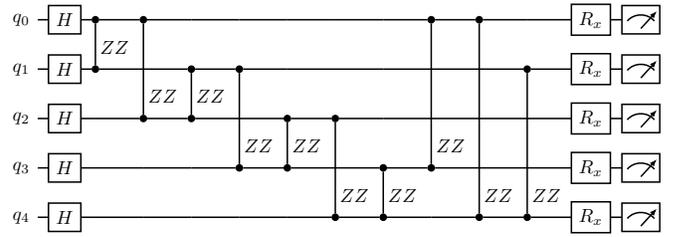

The mathematical description of the cost and mixer functions can be found in Appendix \ref{app:mcp_math}. For the 5 node example graph from Figure \ref{fig:regular_graph}, this yields the circuit shown in Figure \ref{fig:MCP_circuit}.

\subsubsection{Dominating Set Problem}
A dominating set for a graph $G(V,E)$ is defined as the subset $D$ of $V$ such that every member of $D'$ is connected to at least one member of $D$. The dominating set problem aims to find the smallest set $D$ that satisfies this condition. A common allegory is made with a network of surveillance nodes. Each surveillance node can survey itself and each node it is connected to. The question then becomes: what is the smallest number of surveillance nodes needed to survey all nodes in the network?

\begin{figure}[h]
    \centering
    \tikzset{
    phase label/.append style={above right,xshift=-0.1cm,yshift=0.1cm}
    }
    \resizebox{\linewidth}{!}{%
     \begin{quantikz}[column sep = 0.5cm, row sep={0.75cm,between origins}]
    \lstick{$q_0$}           & \gate{H} &\orctrl{1}&\orctrl{1}&\orctrl{1}&\orctrl{1}&\orctrl{1} & \ \ldots\ \qw \\
    \lstick{$q_1$}           & \gate{H} &\orctrl{1}&\orctrl{1}&\orctrl{1}&\orctrl{1}&\orctrl{1} & \ \ldots\ \qw\\
    \lstick{$q_2$}           & \gate{H} &\orctrl{1}&\orctrl{1}&\orctrl{1}&\orctrl{1}&\orctrl{1} & \ \ldots\ \qw\\
    \lstick{$q_3$}           & \gate{H} &\orctrl{1}&\orctrl{1}&\orctrl{1}&\orctrl{1}&\orctrl{1} & \ \ldots\ \qw\\
    \lstick{$q_4$}           & \gate{H} &\orctrl{1}&\orctrl{1}&\orctrl{1}&\orctrl{1}&\orctrl{1} & \ \ldots\ \qw\\
    \lstick{$a_0$}           & \gate{H} &\gate{R_z}&\gate{R_z}&\gate{R_z}&\gate{R_z}&\gate{R_z} & \ \ldots\ \qw\\
    \lstick{$q_0$} \ \ldots\ & \octrl{5}&\qw       &\qw       &\qw       &\qw       &\gate{R_x} & \meter{} \\
    \lstick{$q_1$} \ \ldots\ & \qw      &\octrl{4} &\qw       &\qw       &\qw       &\gate{R_x} & \meter{} \\
    \lstick{$q_2$} \ \ldots\ & \qw      &\qw       &\octrl{3} &\qw       &\qw       &\gate{R_x} & \meter{} \\
    \lstick{$q_3$} \ \ldots\ & \qw      &\qw       &\qw       &\octrl{2} &\qw       &\gate{R_x} & \meter{} \\
    \lstick{$q_4$} \ \ldots\ & \qw      &\qw       &\qw       &\qw       &\octrl{1} &\gate{R_x} & \meter{} \\
    \lstick{$a_0$} \ \ldots\ &\gate{R_z}&\gate{R_z}&\gate{R_z}&\gate{R_z}&\gate{R_z}& \qw       &\qw
    \end{quantikz}%
    }
    \caption{Circuit for the Dominating Set Problem for the 5 node 4-regular graph example. The circuit shows the QAOA implementation for $p=1$. Acilla qubits for the multi-control OR gates (crossed-circles) are left out (4 in total). Notice that controlled $R_z$ gates are inverse-controlled. All $R_z$ and $R_x$ gates depict an $Z$-rotation by $\gamma$ and $X$-rotation by $\beta$ respectively.}
    \label{fig:DSP_circuit}
\end{figure}
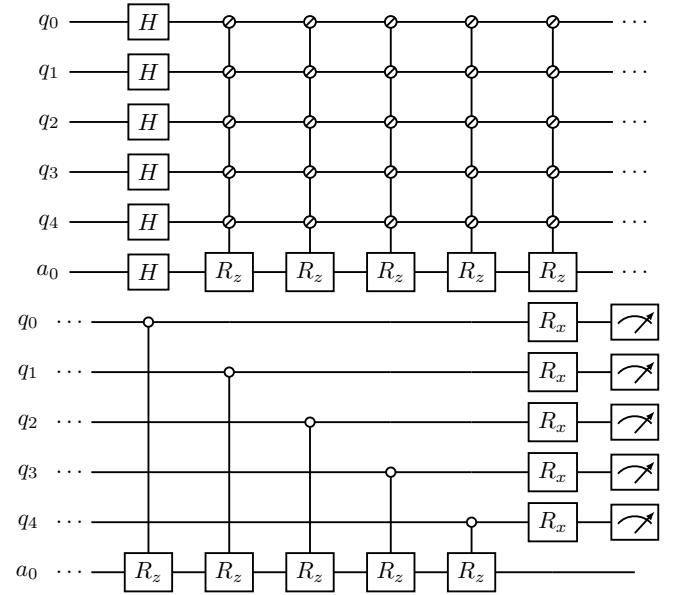

The implementation of the Dominating Set Problem was based on the work by Guerrero~\cite{DSP_QAOA}, using high level multicontrolled quantum OR gates and an ancilla qubit to realize the cost function. Appendix \ref{app:dsp_math} covers the cost function and Appendix \ref{app:gate_OR} describes the multicontrolled quantum OR gate in more detail. Combining this with an initial state generation (all qubits in superposition) and the same mixer unitary as for the MaxCut problem gives us the building blocks for the QAOA circuit. For the example graph, the $p=1$ circuit is shown in Figure \ref{fig:DSP_circuit}.

\subsubsection{Maximum Independent Set Problem}
Consider a graph $G(V,E)$, the set $V' \in V$ is the set of nodes that are not connected to eachother. Find the set of nodes that maximizes the size of $V'$. This problem has a similar feel as the Dominating Set Problem. A solution to the maximum independent set problem is also a solution to the dominating set problem for regular graphs, but not vice versa.

This QAOA implementation  has an interesting mixer (see Appendix \ref{app:mis_math}), coming from a controlled-mixer family in the QAOA generalization~\cite{QAOA_ansatz}. This makes use of the multi-NOR-controlled gates, described in more detail in Appendix \ref{app:mNOR}.

\begin{figure}[h]
    \centering
    \resizebox{\linewidth}{!}{%
     \begin{quantikz}[row sep = 0.1cm]
    \lstick{$q_0$} & \gate{H}  & \gate{R_z} & \gate{R_x} & \norctrl{1}      & \norctrl{1}      & \norctrl{1}      & \norctrl{1}      & \meter{}\\
    \lstick{$q_1$} & \gate{H}              & \gate{R_z} & \norctrl{-1}     & \gate{R_x} & \norctrl{1}      & \norctrl{1}      & \norctrl{1}      & \meter{}\\
    \lstick{$q_2$} & \gate{H}              & \gate{R_z} & \norctrl{-1}     & \norctrl{-1}     & \gate{R_x} & \norctrl{1}      & \norctrl{1}      & \meter{}\\
    \lstick{$q_3$} & \gate{H}              & \gate{R_z} & \norctrl{-1}     & \norctrl{-1}     & \norctrl{-1}     & \gate{R_x} & \norctrl{1}      & \meter{}\\
    \lstick{$q_4$} & \gate{H}              & \gate{R_z} & \norctrl{-1}     & \norctrl{-1}     & \norctrl{-1}     & \norctrl{-1}     & \gate{R_x} & \meter{}
    \end{quantikz}
    }%
    \caption{Maximum Independent Set circuit on the 5 node 4-regular graph example. The circuit shows the QAOA implementation for $p=1$. Ancilla qubits for the multi-control NOR gates (square controls) are left out (3 in total). All $R_z$ and $R_x$ gates depict an $Z$-rotation by $\gamma$ and $X$-rotation by $\beta$ respectively.}
    \label{fig:MIS_circuit}
\end{figure}
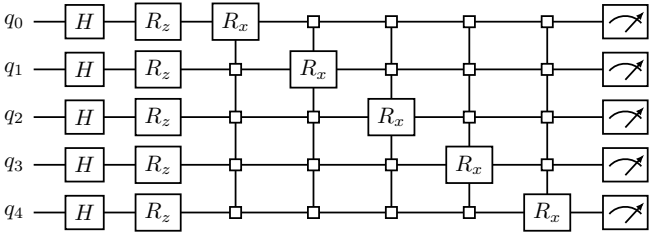

The complete circuit can thus be constructed with a simple cost unitary ($Z$-rotations) and complex mixer unitaries using multi-controlled quantum NOR gates, as shown in in Figure \ref{fig:MIS_circuit}. Ancilla qubits are left out, but it should be noted that three more qubits are used in the decomposed circuit. 

\subsubsection{Travelling Salesperson Problem}
For a graph $G(V,E)$ and distances $d : [V]^2 \rightarrow \mathbb{R}_+$, find an ordering of the nodes that minimizes the total distance traveled for the corresponding tour. A tour visits each node once and returns from the last node to the first~\cite{QAOA_ansatz}. In other words, find the Hamiltonian cycle that with the shortest distance.

The starting point for the implementation of the Traveling Salesperson Problem on a QAOA circuit was presented in a blog by Ceroni~\cite{TSP_QAOA}. In his work, the problem encoding, cost Hamiltonian and Mixer Hamiltonian have been implemented in the QPack TSP algorithm, with some minor tweaks to the original code. An improved state initialization in the form of a Dicke State by Mesman~\cite{QPack_benchmark_paper}, has been implemented for a more uniform distribution of the solution space.

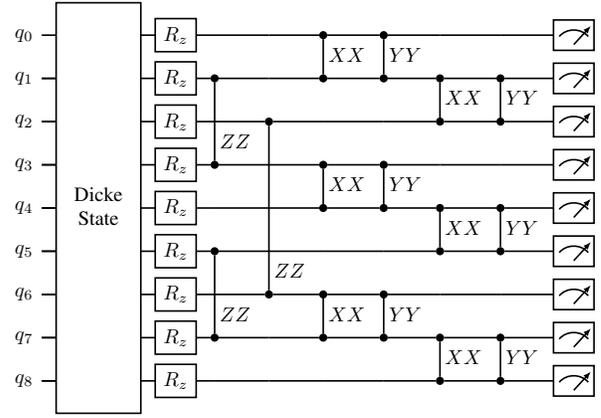
\begin{figure}[h]
    \centering
    \tikzset{
    phase label/.append style={above right,xshift=-0.1cm,yshift=0.1cm}
    }
    \resizebox{0.9\linewidth}{!}{%
     \begin{quantikz}[column sep = 0.25cm, row sep={0.75cm,between origins}]
        \lstick{$q_0$}& \gate[wires=9]{\begin{array}{c} \text{Dicke} \\ \text{State} \end{array}}   &\gate{R_z}&\qw                 &\qw                &\ctrl{1}   &\ctrl{1}   &\qw        &\qw        &\meter{} \\
        \lstick{$q_1$}& \qw                                                 &\gate{R_z}&\ctrl{2}            &\qw                &\rxx{\beta}&\ryy{\beta}&\ctrl{1}   &\ctrl{1}   &\meter{} \\
        \lstick{$q_2$}& \qw                                                 &\gate{R_z}&\qw                 &\ctrl{4}           &\qw        &\qw        &\rxx{\beta}&\ryy{\beta}&\meter{} \\
        \lstick{$q_3$}& \qw                                                 &\gate{R_z}&\rzz{\omega \gamma} &\qw                &\ctrl{1}   &\ctrl{1}   &\qw        &\qw        &\meter{} \\
        \lstick{$q_4$}& \qw                                                 &\gate{R_z}&\qw                 &\qw                &\rxx{\beta}&\ryy{\beta}&\ctrl{1}   &\ctrl{1}   &\meter{} \\
        \lstick{$q_5$}& \qw                                                 &\gate{R_z}&\ctrl{2}            &\qw                &\qw        &\qw        &\rxx{\beta}&\ryy{\beta}&\meter{} \\
        \lstick{$q_6$}& \qw                                                 &\gate{R_z}&\qw                 &\rzz{\omega \gamma}&\ctrl{1}   &\ctrl{1}   &\qw        &\qw        &\meter{} \\
        \lstick{$q_7$}& \qw                                                 &\gate{R_z}&\rzz{\omega\gamma}  &\qw                &\rxx{\beta}&\ryy{\beta}&\ctrl{1}   &\ctrl{1}   &\meter{} \\
        \lstick{$q_8$}& \qw                                                 &\gate{R_z}&\qw                 &\qw                &\qw        &\qw        &\rxx{\beta}&\ryy{\beta}&\meter{} 
    \end{quantikz}%
    }
    \caption{Complete circuit for the Traveling Salesman Example network}
    \label{fig:TSP_circuit}
\end{figure}

An example circuit for a 3 node triangular graph is shown in Figure \ref{fig:TSP_circuit} (using the 5 node example figure would require 25 qubits, which is too large for demonstration purposes). Notice that to model an $n$-node graph, $n^2$ qubits are needed. This is due to the fact that the graphed is encoded to represent its adjacency matrix. This makes the TSP the only QAOA problem currently in QPack that scales quadratic in qubits as the problem size increases.

\subsection{VQE}
\begin{figure}[h]
    \centering
    \includegraphics[width = \linewidth]{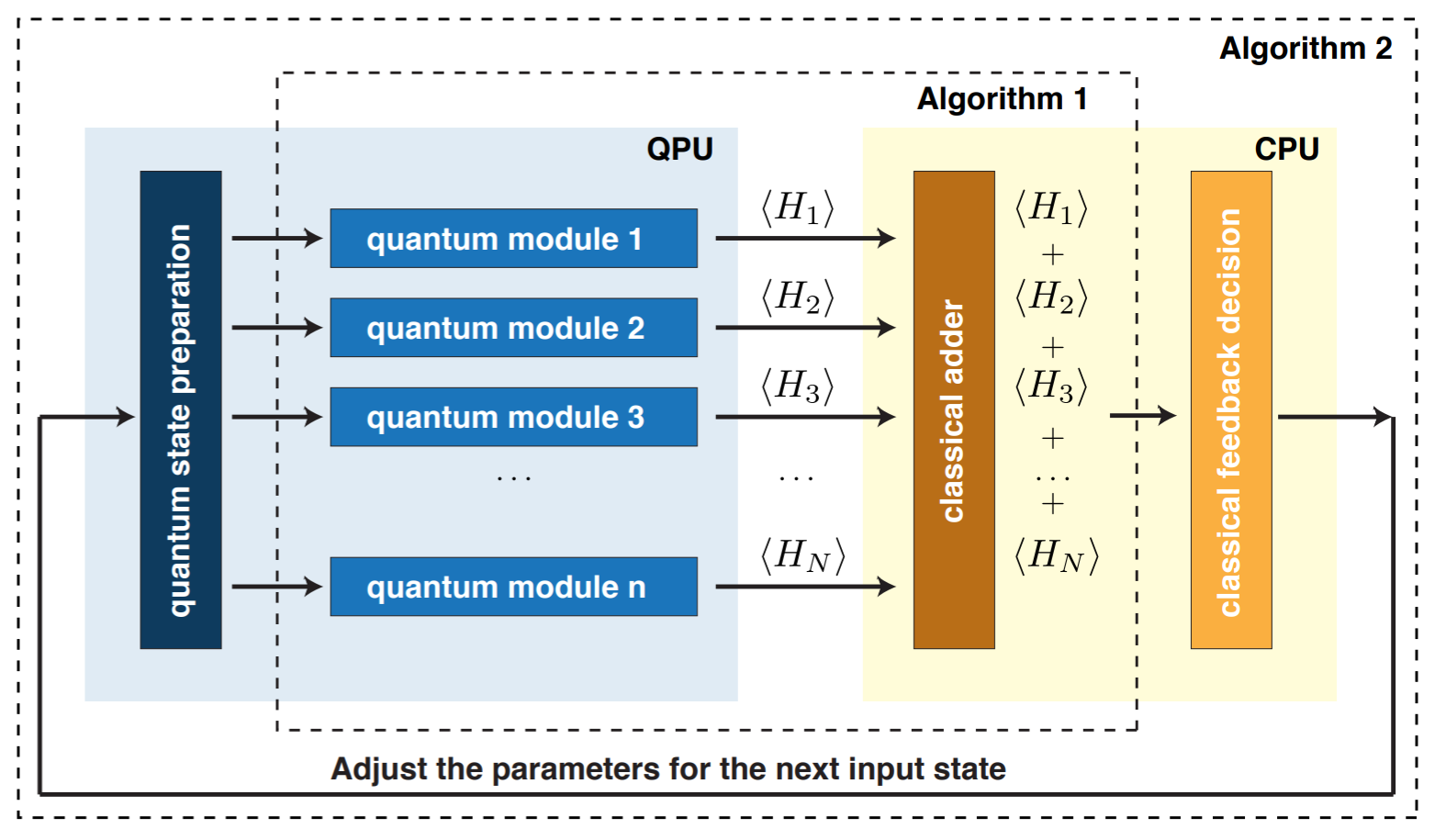}
    \caption{Architecture of the Variational Quantum Eigensolver~\cite{VQE}}
    \label{fig:VQE_overview}
\end{figure}

Like any VQA, the Variational Quantum Eigensolver consists of a classical and quantum subroutine. An overview of the VQE architecture~\cite{VQE} is presented in Figure \ref{fig:VQE_overview}. The first algorithm consists of quantum circuits that compute the expectation value $\langle H_i \rangle$ of Hamiltonian $H = \sum_i^M H_i$. This means that $M$ quantum circuits need to be evaluated to determine the total expectation value $\langle H \rangle$. The partial expectation values are simply added by a classical computer to get the total expectation value. The second algorithm attempts to minimize this expectation value by varying the quantum state $\ket{\psi(\vec{\theta})}$. In QPack, problem sizes equal the number of qubits the Hamiltonian describes, i.e., 1 qubit models 1 particle.

\subsubsection{Random Diagonal Hamiltonian}
This simple implementation shows the basic functionality of the Variational Quantum Eigensolver. For this problem, a diagonal matrix is constructed with random values on the diagonal. The VQE aims to find the lowest eigenvalue of this matrix, which is trivial to find as it is simply the lowest diagonal value.

\begin{equation}
    H_{random} = \sum_{i=0}^{M-1} r_i Z_i
    =
    \begin{pmatrix} 
    R_0    & 0      & 0      & \hdots & 0 \\
    0      & R_1    & 0      & \hdots & 0 \\
    0      & 0      & R_2    & \hdots & 0 \\
    \vdots & \vdots & \vdots & \ddots & \vdots \\
    0      & 0      & 0      & \hdots & R_{M-1}
    \end{pmatrix}
\end{equation}  
\noindent
where $r_i \in [-1,1]$ is a random value, $Z_i$ is the Pauli $Z$ operator on qubit $i$ and $R_i$ is a random value bases on a sum of $r_i$. 

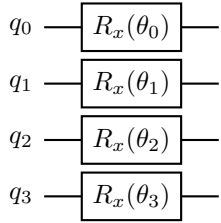
\begin{figure}[h]
    \centering
    \begin{quantikz}[row sep = {0.75cm,between origins}]
        \lstick{$q_0$} & \gate{R_x(\theta_0)} & \qw \\
        \lstick{$q_1$} & \gate{R_x(\theta_1)} & \qw \\
        \lstick{$q_2$} & \gate{R_x(\theta_2)} & \qw \\
        \lstick{$q_3$} & \gate{R_x(\theta_3)} & \qw 
        \end{quantikz}
    \caption{Random Diagonal Hamiltonian Ansatz}
    \label{fig:random_anz}
\end{figure}

The ansatz used to find the lowest eigenvalue is simply an $X$-rotation on each qubit for different optimization parameters, shown in Figure \ref{fig:random_anz}. Since only values on the diagonal of the Hamiltonian are nonzero, this ansatz covers the complete solution space. As the problem size increases, the number of optimization parameters grows proportionally.

\subsubsection{Transverse Ising Chain Model}
This problem focuses on on finding the ground state of a one-dimensional spin system. The model that is used is the 1D Transverse Field Ising Model (TFIM). It consists of spin-1/2 particles on a chain, where neighbors interact with eachother via X-X couplings and a transverse field of strength $h$ is applied in the Z-direction, which alters the spin of the particles in the chain (Equation \ref{eq:ising}). The goal of the VQE algorithm is to find the lowest energy state of this system~\cite{ising_chain} for $h=J$.

\begin{equation} \label{eq:ising}
    H_{Ising} = -J \sum_{i,j} X_i X_j - h\sum_i Z_i
\end{equation}

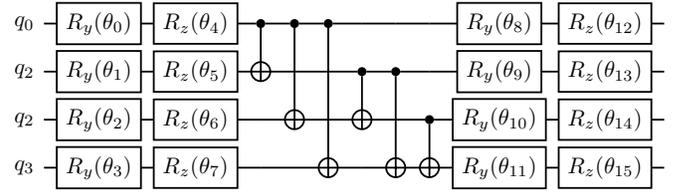
\begin{figure}[h]
    \centering
    \resizebox{\linewidth}{!}{%
    \begin{quantikz}[column sep = 0.2cm, row sep={0.75cm,between origins}]
    \lstick{$q_0$}   & \gate{R_y(\theta_0)} & \gate{R_z(\theta_4)} & \ctrl{1} & \ctrl{2} & \ctrl{3} & \qw      & \qw      & \qw      & \gate{R_y(\theta_8)} & \gate{R_z(\theta_{12})} & \qw \\
    \lstick{$q_2$}   & \gate{R_y(\theta_1)} & \gate{R_z(\theta_5)} & \targ{}  & \qw      & \qw      & \ctrl{1} & \ctrl{2} & \qw      & \gate{R_y(\theta_9)} & \gate{R_z(\theta_{13})} & \qw \\
    \lstick{$q_2$}   & \gate{R_y(\theta_2)} & \gate{R_z(\theta_6)} & \qw      & \targ{}  & \qw      & \targ{}  & \qw      & \ctrl{1} & \gate{R_y(\theta_{10})} & \gate{R_z(\theta_{14})} & \qw \\
    \lstick{$q_3$}   & \gate{R_y(\theta_3)} & \gate{R_z(\theta_7)} & \qw      & \qw      & \targ{}  & \qw      & \targ{}  & \targ{}  & \gate{R_y(\theta_{11})} & \gate{R_z(\theta_{15})} & \qw \\
    \end{quantikz}
    }%
    \caption{Efficient SU2 ansatz~\cite{SU2_ansatz} with full entanglement for 4 qubits}
    \label{fig:su2}
\end{figure}

Ansatzes for this problem can become very hardware demanding as the problem size grows. To negate this, the hardware efficient SU2~\cite{SU2_ansatz} will be used with full entanglement, see Figure \ref{fig:su2}.  This circuit explores the solution space for the Hamiltonian using only a few gates, but also needs more optimization parameters then the simply Random Hamiltonian ansatz, 4 for every qubit. 

\subsection{Improved Hamiltonian measurement approach}
To reduce the number of circuits needed to evaluate the total expectation value $\langle H \rangle$, the bases that form the Hamiltonian are analyzed. Notice that for the implemented VQE problems, only $Z$ or $XX$ basis need to be evaluated. Instead of measuring each individual Hamiltonian component, all qubits can be measured in the $Z$ or $X$ basis, which have the partial expectation values of $\langle H_i \rangle$ encoded. This reduces the number of $M$ circuits to be evaluated to one and two for the Random Hamiltonian and Ising Chain problem respectively, and therefore the total execution time of the QPack benchmark while still gathering enough quantum execution data for performance evaluation. Thus, instead of needing $M$ circuits to evaluate, only one is needed for the Random Hamiltonian problem and two are needed for the Ising Chain problem.

\section{Qpack overview} \label{sec:qpack_overview}

Building on previous work~\cite{QPack_benchmark_paper}, this version of QPack is implemented in the LibKet library and has extended the number of quantum algorithms used for benchmarking. This section briefly describes the structure of QPack in its current form.

\subsection{LibKet}
In order for the benchmark to be more versatile, a cross-platform library for quantum computers is desirable, as a quantum circuit can be build as a general quantum expression and executed on multiple quantum devices. This will not only allow for efficient access to current quantum devices, but for easy integration with future devices as well.

In 2018, TU Delft presented a new cross-platform programming framework called $\ket{\text{Lib}}$ (LibKet)~\cite{libket}. This framework uses a CUDA inspired streaming concept to execute quantum hardware as an accelerator, like one would with a GPU. LibKet is designed based on the principles of quantum-accelerated computing, concurrent task offloading, single-source quantum-classical programming, using generic expressions, development on top of existing tools and seamless integration into the status quo. Programs are written natively in C++, but some early stage Python and C APIs are developed as well. According to the online documentation~\cite{libket_docs}, LibKet supports both remote quantum devices as well as local simulators, such as those provided by Qiskit \cite{qiskit}, IBMQ~\cite{IBMQ}, Cirq~\cite{cirq}, PyQuil~\cite{pyquil} and QuEST~\cite{quest}.

Using the above-mentioned capabilities of LibKet, it will serve as a basis for the QPack benchmark, allowing for an efficient method of measuring different quantum computers using a single quantum program. Other frameworks such as XACC~\cite{xacc, xacc_docs} or SuperstaQ~\cite{superstaq} could also be viable platforms to create cross-platform benchmarks, but LibKet was chosen due to its customizability.

\subsection{Benchmark outline} \label{subsec:outline}
The QPack benchmark determines the performance of quantum computers by collecting and evaluating quantum execution data when executing various VQAs. These VQAs are one of the first viable real-use applications in the NISQ-era of quantum computing~\cite{VQA} and have therefore been chosen as applications to run in this benchmark. Figure \ref{fig:qpack_flowchart} shows an overview of the QPack benchmarking process.\\

\tikzstyle{decision} = [diamond, draw, fill=green!20, text width=5em, text badly centered, inner sep=0pt]
\tikzstyle{block} = [rectangle, draw, fill=blue!20, text width=5em, text centered, rounded corners, minimum height=4em]
\tikzstyle{line} = [draw, -latex']
\tikzstyle{cloud} = [ellipse, draw, fill=red!20, text width=7em, text centered, minimum height=2em]

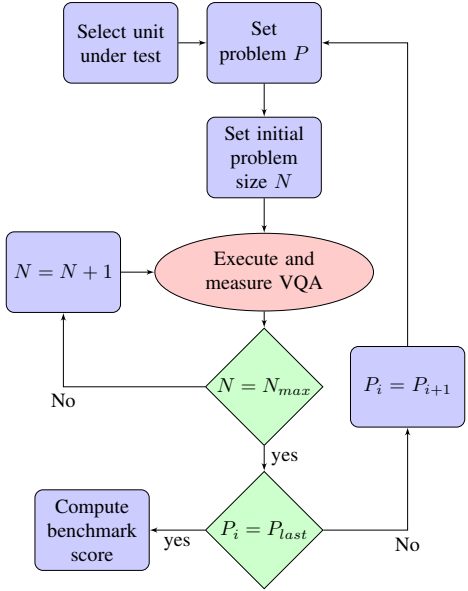
\begin{figure}[h]
    \centering
    \resizebox{0.7\linewidth}{!}{%
    \begin{tikzpicture}[node distance = 2cm, auto]
    \node [block] (QPU) {Select unit under test};
    \node [block, right of=QPU, node distance = 2.5cm] (Problem) {Set problem $P$};
    \node [block, below of=Problem] (Size) {Set initial problem size $N$};
    \node [cloud, below of=Size] (Execute) {Execute and measure VQA};
    \node [block, left of=Execute, node distance=3.5cm] (incN) {$N=N+1$};
    \node [decision, below of=Execute, node distance=2cm] (maxN) {$N=N_{max}$};
    \node [block, right of=maxN, node distance=2.5cm] (incP) {$P_i=P_{i+1}$};
    \node [decision, below of=maxN, node distance=2.5cm] (maxP) {$P_i=P_{last}$};
    \node [block, left of=maxP, node distance=3cm] (Score) {Compute benchmark score};

    \path [line] (QPU) -- (Problem);
    \path [line] (Problem) -- (Size);
    \path [line] (Size) -- (Execute);
    \path [line] (Execute) -- (maxN);
    \path [line] (maxN) -| node [below] {No} (incN);
    \path [line] (incN) -- (Execute);
    \path [line] (maxP) -| node [below] {No} (incP);
    \path [line] (incP) |- (Problem);
    \path [line] (maxN) -- node {yes} (maxP);
    \path [line] (maxP) -- node [below]{yes} (Score);
\end{tikzpicture}
    }%
    \caption{Benchmark process overview}
    \label{fig:qpack_flowchart}
\end{figure}  

The benchmark starts by selecting a quantum device to be evaluated. Multiple devices can be selected for sequential performance evaluation, but this overview will just focus on the evaluation of a single quantum computer. After device selection, a problem from the problem set $P$ can be selected. In the current implementation of QPack, $P \in \{\text{MCP, DSP, MIS, TSP, RH, IC}\}$ as described in Appendix \ref{sec:algos}. This set contains four QAOA problems (MaxCut, Dominating Set, Maximum Independent Set and Traveling Salesperson problems) and 2 VQE problems (Random diagonal Hamiltonian and Ising Chain model) respectively. There is no preferred order, so a problem set can be selected freely. However, in order to get the best and most varied comparison, QPack attempts to complete all problems for as large a problem size as possible. \\ 

For the selected problem, the initial problem size is set. Problem size can differ by the number of qubits used for a given problem, hence a selection of problem sizes should be carefully considered. For example, the MaxCut's problem scales qubits linearly with the problem size (problem size 5 needs 5 qubits), while the Traveling Salesperson Problem scales quadratic (problem size 4 requires 16 qubits). For this reason, each problem has its own range of problem sizes.\\

When the problem and problem size is set, the VQA can be run on the quantum computer under test. This entails the optimization of the parameterized quantum circuit by a classical optimizer (COBYLA~\cite{COBYLA} is used in this version of QPack). Once the optimizer has found the optimal minimal value, the VQA execution is finished and the measurement results are saved for use later. This step is repeated 10 times, such that measurements are taken in multitude. Specific details on what measurement data is being collected are elaborated in Section \ref{sec:metrics}.\\

After VQA execution is finished, QPack checks if the maximum problem size has been evaluated. If not, the problem size is incremented and the VQA is measured for this new problem size. When all problem in the range have finished, QPack checks if all problems have been evaluated. If not, the next problem is set up and the VQA evaluation is repeated for this new problem. When all problems have finished, QPack evaluates all measured data and computes the benchmark score.\\

\section{QAOA cost functions and mixers} \label{app:cost_mix}

\subsection{MaxCut Problem} \label{app:mcp_math}
\paragraph{Encoding}
The nodes can be encoded in a single bitstring $x= x_1 x_2 ... x_V$ for $V$ nodes. If a node is in the first distinct set, its bit will be 0 and if it is in the other set, its bit will be 1. The QAOA will therefore require $V$ qubits to encode the solution space of the MaxCut problem.

\paragraph{Initial State}
Since all possible permutations of bitstring $x$ form valid potential solutions to the MaxCut problem, the initial state can be initialized to be an equal superposition of all states. Hence, the initial state $\ket{s}$ is:
\begin{equation}
    \ket{s} = \prod_{i \in V} \ket{+}_i
\end{equation}

\paragraph{Cost Hamiltonian}
The cost function will be the number of cuts a bitstring can make. So for an edge set $\{u,v\}$ of size $E$. The cost Hamiltonian is

\begin{equation}
    H_C = \sum_{\{u,v\}} \frac{1}{2}(1-\sigma_u^z \sigma_v^z)
\end{equation}

Since the global phase term can be ignored, the cost Hamiltonian simplifies to
\begin{equation}
    H_C = \sum_{\{u,v\}} \sigma_u^z \sigma_v^z
\end{equation}
\noindent
and the cost unitary is
\begin{equation}
    U_C(\gamma) = e^{-i\gamma H_C} = \prod_{\{u,v\}} e^{-i\gamma \sigma_u^z \sigma_v^z}
\end{equation}

\paragraph{Mixer Hamiltonian}
In order to mix up the solution space, simple Pauli-x gates can be applied to all qubits, as expressed by Hamiltonian
\begin{equation}
    H_M = \sum_{i \in V} \sigma_i^x
\end{equation}
\noindent
resulting in the unitary
\begin{equation}
    U_M(\beta) = e^{-i\beta H_M} = \prod_{i \in V} e^{-i\beta\sigma_i^x}
\end{equation}
\noindent
which can simply be implemented with an $R_x(\beta)$ gate on each qubit.

\subsection{Dominating Set Problem} \label{app:dsp_math}
An implementation for the Dominating Set problem using QAOA was presented by Guerrero in 2020~\cite{DSP_QAOA}. The solution makes use of the high level quantum OR-gate to encode the cost function using an ancilla qubit. For a complete graph, this approach requires $2V$ qubits. However, QPack uses the 4-regular graph as the problem input, this is reduced to only needing $V+5$ qubits, which will become clear later this section as the cost Hamiltonian is elaborated.

\paragraph{Encoding}
The solution is encoded in a bitstring $x = x_1 x_2 ... x_V$. If a bit is in the subset, $D$ is is 1, otherwise it is 0. All possible combinations of the bitstring $x$ are valid (but not necessarily correct) solutions to the DSP.

\paragraph{Initial state}
Similar to the MaxCut problem, all possible combinations of bitstring $x$ form valid potential solutions to the Dominating Set problem. Thus, the initial state can be initialized to be an equal superposition of all states. Hence, the initial state $\ket{s}$ is:
\begin{equation}
    \ket{s} = \prod_{i \in V} \ket{+}_i
\end{equation}

\paragraph{Cost Hamiltonian}
Guerrero splits the cost function into two clauses, $T_k(x)$ and $D_k(x)$. The first clause measures the number of clauses surveyed and is defined as:
\begin{equation}
    T_k(x)= \begin{cases}
1 &\text{if $x_k$ is connected to any $x_i = 1$ }\\
0 &\text{otherwise}
\end{cases}
\end{equation}

The second clause measures the number of surveillance nodes used, i.e., the number of ones in the string $x$:

\begin{equation}
    D_k(x)= \begin{cases}
1 &\text{if $x_k = 0$ }\\
0 &\text{if $x_k = 1$ }
\end{cases}
\end{equation}

Both clauses will be encoded onto an ancilla qubit that sums the cost. Implementing clauses like this will require an extra qubit along the $V$ qubits that are necessary to encode the solution space.

For the implementation of the $T_k(x)$ clause, a multi-OR-controlled quantum $R_z(\theta)$ gate is required (see Appendix \ref{app:mOR}). For every node in the graph $G(V,E)$, such a gate is required, with control qubits being the node itself and its neighbors, targeting the cost ancilla qubit. Since for a regular graph, a node always has 4 neighbors, a maximum of 5-OR-controlled $R_z(\theta)$ gate is required, which uses 4 extra ancilla qubits. This makes the totally needed qubits of the Dominating Set Circuit equal to $V+5$ for a 4-regular graph.\\

The $D_k(x)$ clause is simpler to implement, which can be achieved by a bit flip on every bit in the bitstring. This is done by an inverted controlled $R_z(\theta)$ gates on the cost ancilla qubit. This gate can simply be implemented by using two $X$-gates and a controlled $R_z(\theta)$ gate.

\paragraph{Mixer Hamiltonian}
Similar to the MaxCut problem, the mixer Hamiltonian  can simply be implemented with an $R_x(\beta)$ gate on each qubit that encodes the $x$ bitstring. 

\subsection{Maximal Independent Set Problem} \label{app:mis_math}
\paragraph{Encoding}
The nodes are encoded onto a bitstring $\Vec{x} = x_0 x_1 ... x_V$, where a node is in $V$ if $x_i = 1$, otherwise $x_i  = 0$. 

\paragraph{Initial state}
The initial qubit state can be any arbitrary state since all possible states are possible solutions to the MIS problem for any graph. To stay consistent with MCP and DSP, all qubits are initialized in maximum superposition as well.

\begin{equation}
    \ket{s} = \prod_{i \in V} \ket{+}_i
\end{equation}

\paragraph{Cost Hamiltonian}
The cost function is the size of the independent set, which is simply the sum of all bits $x_i$ in the bitstring: $\sum_{i=1}^V x_i$. Using a transformation of the objective function~\cite{QAOA_ansatz}, this function can be translated to Hamiltonian

\begin{equation}
    H_C = \sum_{i=1}^V Z_i
\end{equation}
\noindent
which can be implemented with a depth-1 circuit of Z-rotations on each qubit.

\paragraph{Mixer Hamiltonian}
The Maximal Independent Set QAOA mapping provides a more complex family of mixers compared to the MaxCut or Dominating Set problems. This is a set of controlled-mixer operations. For the MIS problem, we only want to mix the solution space (i.e., adding or removing nodes from the set $'V$) based on its neighboring nodes and whether they are already in the solution space or not.  The mixer Hamiltonian is thus given by

\begin{equation}
    H_M,i = X_i H_{NOR(x_{neighbor(i)})}
\end{equation}
\noindent
To build this Hamiltonian, a set of partitioned controlled mixers is used~\cite{QAOA_ansatz}, where 

\begin{equation}
    U_M,i(\beta) = \Lambda_{NOR(x_{neighbor(i)})}(e^{-i\beta X_i})
\end{equation}
\noindent
where the $\Lambda_{NOR}$ function denotes the control NOR operator. This is implemented using a quantum NOR-RotateX operation, described in Appendix \ref{app:mNOR}. It becomes clear that this circuit needs an extra ancilla qubit, as well as two more ancilla qubits for the multi-controlled quantum NOR gates since the maximum number of neighbors in the 4-regular graph is 4. This requires the MIS circuit to have V+3 qubits.

\subsection{Traveling Salesperson Problem} \label{app:tsp_math}

The starting point for the implementation of the Traveling Salesman Problem was presented in a blog by Ceroni~\cite{TSP_QAOA}. In his work, the problem encoding, cost Hamiltonian and mixer Hamiltonian have been implemented in the QPack TSP algorithm, with some minor tweaks to the original code. An improved state initialization has been implemented for a more uniform distribution of the solution space.

\paragraph{Encoding}
In order to get a path, the graph will be represented in an adjacency matrix and a corresponding distance matrix, which contains the distances of each node to each other.

\begin{equation*}
    A = \begin{pmatrix} 0&1&1\\1&0&1\\1&1&0 \end{pmatrix}, \hspace{1cm}  D = \begin{pmatrix} 10&1&1\\1&10&1\\1&1&10 \end{pmatrix}
\end{equation*}
\noindent
where a node itself has weight 10, to discourage the path to travel to itself. The adjacency matrix is then mapped onto a qubit string by the matrix index. For instance, the example adjacency matrix can be encoded onto the string $x = 011101110$. 

\paragraph{Initial state}
Since not all string values are a valid solution to the problem (e.g., only strings that encode a valid adjacency matrix), initializing all qubits in uniform superposition over all states would not be efficient. Instead, all qubits encoding a row of the adjacency matrix are encoded into a superposition of states with Hamming weight 2, known as Dicke State $\ket{D^n_k}$~\cite{Dicke_def}, defined as the equal superposition of all $n$-qubit states $\ket{x}$ with Hamming weight $wt(x) = k$ (see Appendix \ref{app:dicke}). For a 3-node triangular network, this would require the initial state to be:

\begin{equation}
    \ket{D^3_2} = \frac{1}{\sqrt{3}}(\ket{011} + \ket{101} + \ket{110})
\end{equation}

\paragraph{Cost Hamiltonian}

The cost Hamiltonian for the Traveling Salesman Problem consists of two parts. A soft constraint and hard constraint. The soft constraint is defined as follows:

\begin{equation}
    H_{soft} = \sum_{i,j} \frac{D_{ij}}{2}(1-Z_{ij})
\end{equation}
\noindent
where $D_{ij}$ is the distance between two nodes $i$ and $j$, defined by the distance matrix. Dropping the global phase term results in

\begin{equation}
    H_{soft} = -\frac{1}{2}\sum_{i,j} D_{ij}Z_{ij}
\end{equation}

This constraint will force the result to an adjacency matrix for the shortest path solution. The hard constraint is defined as follows:

\begin{equation}
    H_{hard} = -\sum{a,b}Z_a Z_b
\end{equation}
\noindent
where $a=A_{ij}$ and $b=A_{ji}$. This results in ZZ-rotations for every two qubit pairs that represent the reflection across the diagonal of the adjacency matrix. This essentially forces the resulting bitstring to produce a symmetric matrix. The final cost Hamiltonian is then

\begin{equation}
    H_C = H_{soft} + \omega H_{hard} = -\frac{1}{2}\sum_{i,j} D_{ij}Z_{ij} - \omega \sum{a,b}Z_a Z_b
\end{equation}
\noindent
where $\omega$ must be chosen such that it is relatively large to the soft constraint Hamiltonian. According to~\cite{TSP_QAOA}, $\omega = 5$ is sufficient.

\paragraph{Mixer Hamiltonian}
Finally the solution space can be mixed with the mixer Hamiltonian, which can be achieved with SWAP gates between the adjacent qubits of a row. The SWAP operations conserve the Hamming weight and thus the solution space is preserved. The mixer Hamiltonian then is

\begin{equation}
    H_M = \sum_{i,j} SWAP_{i,j} = \frac{1}{2}\sum_{i,j}(X_i X_j + Y_i Y_j)
\end{equation}

Which is an XX- and YY-rotation for each adjacent qubits pair encoding a row.
\section{High-level gates} \label{app:gates}

\subsection{Quantum OR} \label{app:gate_OR}
\begin{figure}[h]
    \centering
     \begin{quantikz}
     &  \orctrl{1} & \qw \midstick[3,brackets=none]{=}  & \ctrl{1} & \ctrl{2} & \qw       & \qw \\
     &  \orctrl{1} & \qw                                & \ctrl{1} & \qw      & \ctrl{1}  & \qw \\
     &  \targ{}    & \qw                                & \targ{}  & \targ{}  & \targ{}   & \qw
    \end{quantikz}
    \caption{Quantum OR gate decomposition}
    \label{fig:QOR}
\end{figure}
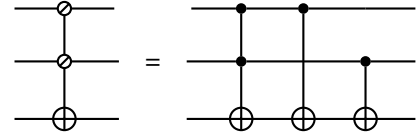

A three qubit quantum OR gate can be decomposed into a Toffoli gate and two CNOT gates, shown in figure \ref{fig:QOR}.

\subsection{Multicontrolled gates} \label{app:gate_multi}

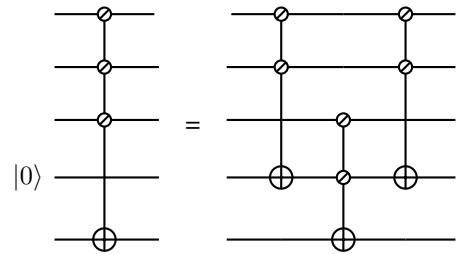
\begin{figure}[h]
    \centering
     \begin{quantikz}
    \qw                &  \orctrl{1} & \qw \midstick[5,brackets=none]{=}  & \orctrl{1} & \qw        & \orctrl{1} & \qw \\
    \qw                &  \orctrl{1} & \qw                                & \orctrl{2} & \qw        & \orctrl{2} & \qw \\
    \qw                &  \orctrl{2} & \qw                                & \qw        & \orctrl{1} & \qw        & \qw \\
    \lstick{$\ket{0}$} & \qw         & \qw                                & \targ{}    & \orctrl{1} & \targ{}    & \qw \\
     \qw               &  \targ{}    & \qw                                & \qw        & \targ{}    & \qw        & \qw
    \end{quantikz}
    \caption{Multi-control Quantum OR gate decomposition}
    \label{fig:mQOR}
\end{figure}

The multicontrolled Quantum OR gate is then simply a combination of three qubit Quantum OR gates. For every extra control qubit, an extra ancilla qubit is needed as well. For example, a 3-control OR gate is decomposed as shown in Figure \ref{fig:mQOR}.

\subsection{OR controlled gates} \label{app:mOR}
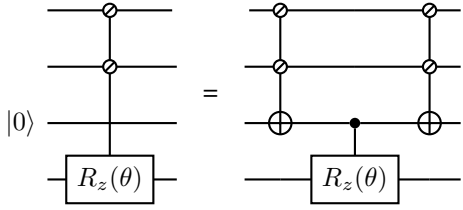
\begin{figure}[h]
    \centering
     \begin{quantikz}[column sep = 0.25cm, row sep={0.75cm,between origins}]
    \qw                &  \orctrl{1} & \qw \midstick[5,brackets=none]{=}  & \orctrl{1} & \qw        & \orctrl{1} & \qw \\
    \qw                &  \orctrl{2} & \qw                                & \orctrl{1} & \qw        & \orctrl{1} & \qw \\
    \lstick{$\ket{0}$} & \qw         & \qw                                & \targ{}    & \ctrl{1} & \targ{}    & \qw \\
     \qw               &  \gate{R_z(\theta)}    & \qw                     & \qw        & \gate{R_z(\theta)}     & \qw        & \qw
    \end{quantikz}
    \caption{Multi-OR-control $R_z(\theta)$ gate decomposition}
    \label{fig:QORRZ}
\end{figure}

Any OR-controlled gate can be decomposed into two multi-control Quantum OR gates and the desired gate, as shown in Figure \ref{fig:QORRZ} with an $Rz(\theta)$ gate as example.

\subsection{NOR-controlled gates} \label{app:mNOR}
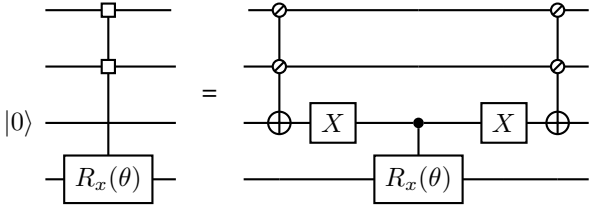
\begin{figure}[h]
    \centering
     \begin{quantikz}[column sep = 0.25cm, row sep={0.75cm,between origins}]
    \qw                & \norctrl{1}        & \qw \midstick[5,brackets=none]{=} & \orctrl{1} & \qw      & \qw               & \qw      & \orctrl{1} & \qw \\
    \qw                & \norctrl{2}        & \qw                               & \orctrl{1} & \qw      & \qw               & \qw      & \orctrl{1} & \qw \\
    \lstick{$\ket{0}$} & \qw               & \qw                                & \targ{}    & \gate{X} & \ctrl{1}          & \gate{X} & \targ{}    & \qw \\
    \qw                & \gate{R_x(\theta)} & \qw                                & \qw        & \qw      & \gate{R_x(\theta)} & \qw      & \qw        & \qw 
    \end{quantikz}
    \caption{Multi-OR-control $R_z(\theta)$ gate decomposition}
    \label{fig:MIS_HM}
\end{figure}

This is implemented using a quantum NOR-RotateX operation, described in Figure \ref{fig:MIS_HM}. This is the Quantum OR circuit as elaborated in the Dominating Set Problem section, but with an added two X gates that function as a NOT gate.

\subsection{XX-, YY- and ZZ-rotations} \label{app:gate_double_paulis} 
The double Pauli rotations encoded in some problem Hamiltonians can be constructed by using two CNOT gates and an $R_z$-gate. To create the $X$, $Y$ or $Z$ rotation, the qubit has to be changed to the desired basis. Thus, for the $XX$-rotation, two Hadamard gates are applied before and after the CNOT gates to change to the $X$-basis and then revert back to the original basis, see Figure \ref{fig:RXX}.

\def\rxx#1{\phase{XX(#1)}}
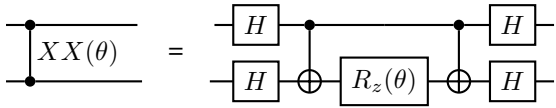
\begin{figure}[h]
    \centering
    \tikzset{
    phase label/.append style={above right,xshift=-0.1cm,yshift=0.0cm}
    }    
     \begin{quantikz}[column sep = 0.25cm, row sep={0.75cm,between origins}]
     &  \ctrl{1}     & \qw \midstick[2,brackets=none]{=}  & \gate{H} & \ctrl{1} & \qw                & \ctrl{1} & \gate{H} & \qw \\
     &  \rxx{\theta} & \qw                                & \gate{H} & \targ{}  & \gate{R_z(\theta)} & \targ{}  & \gate{H} & \qw
    \end{quantikz}
    \caption{$R_{xx}(\theta)$ gate decomposition}
    \label{fig:RXX}
\end{figure}

Similar to the $YY$-rotation, the basis change to the $Y$-basis can be achieved with an $R_x(-\frac{\pi}{2})$ gate and its conjugate transpose, as shown in Figure \ref{fig:RYY}.

\def\ryy#1{\phase{YY(#1)}}
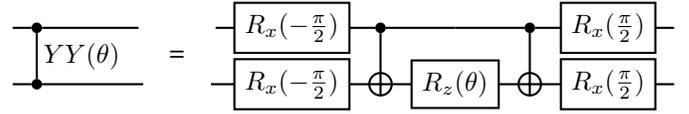
\begin{figure}[h]
    \centering
    \tikzset{
    phase label/.append style={above right,xshift=-0.1cm,yshift=0.0cm}
    }    
     \begin{quantikz}[column sep = 0.25cm, row sep={0.75cm,between origins}]
     &  \ctrl{1}     & \qw \midstick[2,brackets=none]{=}  & \gate{R_x(-\frac{\pi}{2})} & \ctrl{1} & \qw                & \ctrl{1} & \gate{R_x(\frac{\pi}{2})} & \qw \\
     &  \ryy{\theta} & \qw                                & \gate{R_x(-\frac{\pi}{2})} & \targ{}  & \gate{R_z(\theta)} & \targ{}  & \gate{R_x(\frac{\pi}{2})} & \qw
    \end{quantikz}
    \caption{$R_{yy}(\theta)$ gate decomposition}
    \label{fig:RYY}
\end{figure}

The $ZZ$-gate does not require a basis change, see Figure \ref{fig:RZZ}.

\def\rzz#1{\phase{ZZ(#1)}}
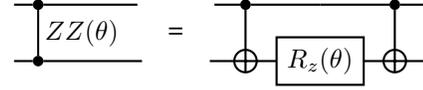
\begin{figure}[h]
    \centering
    \tikzset{
    phase label/.append style={above right,xshift=-0.1cm,yshift=0.0cm}
    }    
     \begin{quantikz}[column sep = 0.25cm, row sep={0.75cm,between origins}]
     &  \ctrl{1}  & \qw \midstick[2,brackets=none]{=}  & \ctrl{1} & \qw & \ctrl{1} & \qw \\
     &  \rzz{\theta}    & \qw     & \targ{}  & \gate{R_z(\theta)} & \targ{}  & \qw
    \end{quantikz}
    \caption{$R_{zz}(\theta)$ gate decomposition}
    \label{fig:RZZ}
\end{figure}

\subsection{Dicke state} \label{app:dicke}
Since not all string values are a valid solution to the problem (e.g. only strings that encode a valid adjacency matrix), initializing all qubits in uniform superposition over all states would not be efficient. Instead, all qubits encoding a row of the adjacency matrix are encoded into a superposition of states with Hamming weight 2, known as Dicke State $\ket{D^n_k}$~\cite{Dicke_def}, defined as the equal superposition of all $n$-qubit states $\ket{x}$ with Hamming weight $wt(x) = k$. For example, in a 3-node triangular graph, a row of its adjacency matrix would be encoded as:

\begin{equation}
    \ket{D^3_2} = \frac{1}{\sqrt{3}}(\ket{011} + \ket{101} + \ket{110})
\end{equation}

The quantum circuit to encode the row this way is presented in~\cite{Dicke_Prep}. The resulting initialization circuit for each row of the TSP example (see Appendix \ref{sec:algos}) graph will thus be:

\begin{figure}[h]
    \centering
     \begin{quantikz}[column sep = 0.25cm, row sep={0.75cm,between origins}]
     \lstick{$q_0$} & \qw      & \qw      & \qw                       & \qw      & \ctrl{2} & \gate{\sqrt{\frac{2}{3}}} & \ctrl{2} & \ctrl{1} & \gate{\sqrt{\frac{1}{2}}} & \ctrl{1} & \qw \\
     \lstick{$q_1$} & \gate{X} & \ctrl{1} & \gate{\sqrt{\frac{1}{3}}} & \ctrl{1} & \qw      & \ctrl{-1}                 & \qw      & \targ{}  & \ctrl{-1}                 & \targ{}  & \qw \\
     \lstick{$q_2$} & \gate{X} & \targ{}  & \ctrl{-1}                 & \targ{}  & \targ{}  & \ctrl{-1}                 & \targ{}  & \qw      & \qw                       & \qw      & \qw
    \end{quantikz}
    \caption{Row initialisation for Travelling Salesman problem}
    \label{fig:DickeInit}
\end{figure}
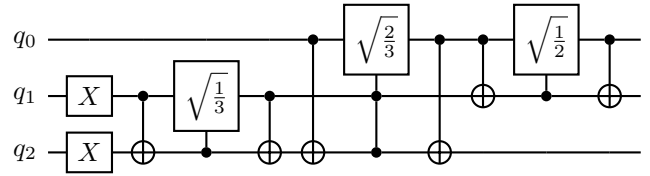

Where the controlled controlled square root $\sqrt{\frac{l}{n}}$ denotes a controlled $R_y(2 acos(\sqrt{\frac{l}{n}}))$ gate.
\section{Quantum computer datasets} \label{app:datasets}
This appendix section lists the quantum execution data of the tested local simulators, remote simulators and remote hardware. For all devices, the quantum and classical runtime are plotted (both average job durations and total job durations) as well as some other data such as quantum layer operations per second, optimizer expectation value, optimizer error compared to the QuEST baseline function and classical optimizer iterations. 

\newcommand\mywidthx{\linewidth}

\subsection{Local Simulators}

\begin{figure}[h!]
    \centering
    \includegraphics[width = \mywidthx]{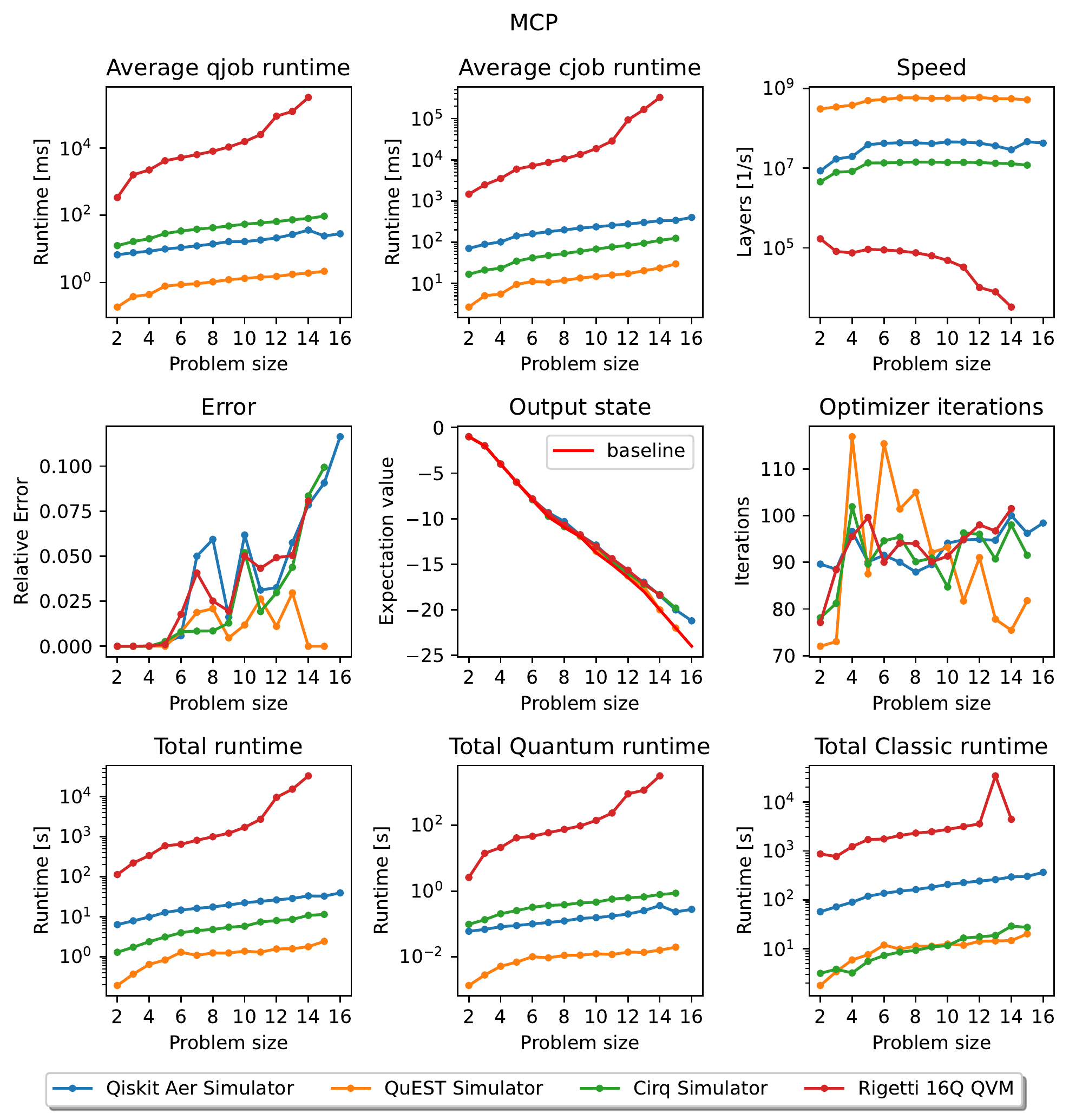}
    \caption{Benchmark results for the MaxCut Problem}
    \label{fig:app_loc_sim_mcp}
\end{figure}

\begin{figure}[h!]
    \centering
    \includegraphics[width = \mywidthx]{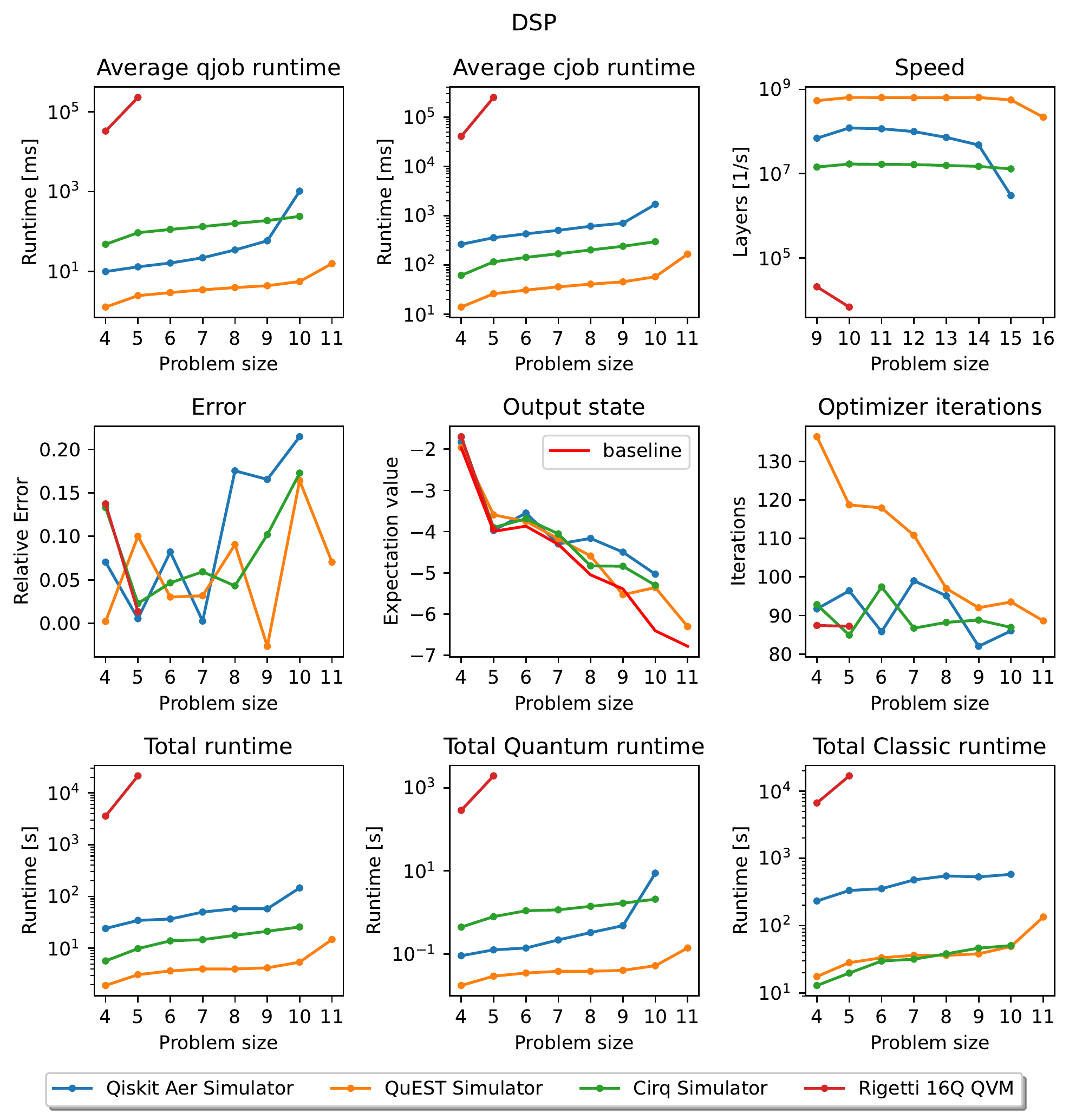}
    \caption{Benchmark results for the Dominating Set Problem}
    \label{fig:app_loc_sim_dsp}
\end{figure}

\begin{figure}[h!]
    \centering
    \includegraphics[width = \mywidthx]{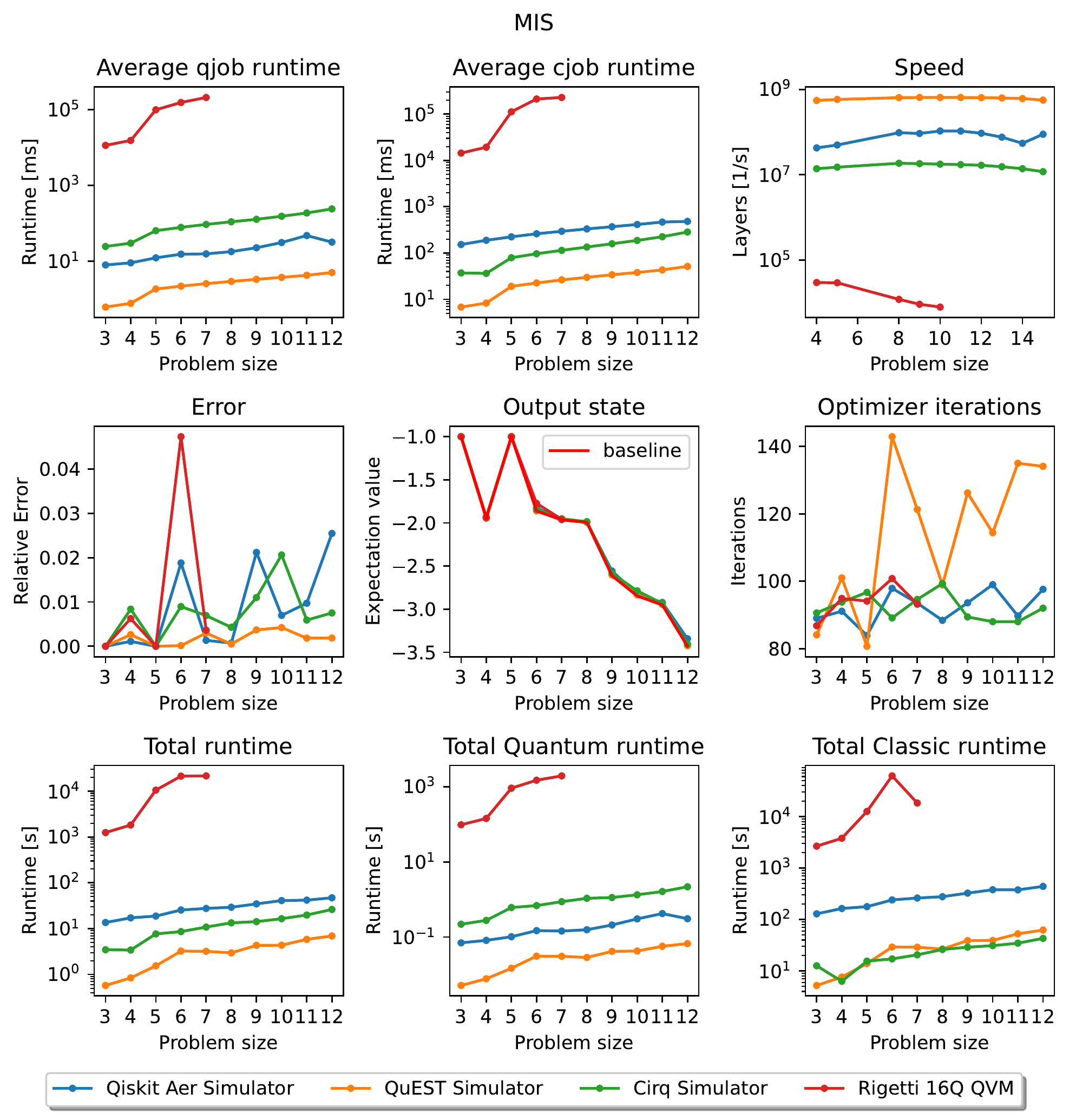}
    \caption{Benchmark results for the Maximum Independent Set Problem}
    \label{fig:app_loc_sim_mis}
\end{figure}

\begin{figure}[h!]
    \centering
    \includegraphics[width = \mywidthx]{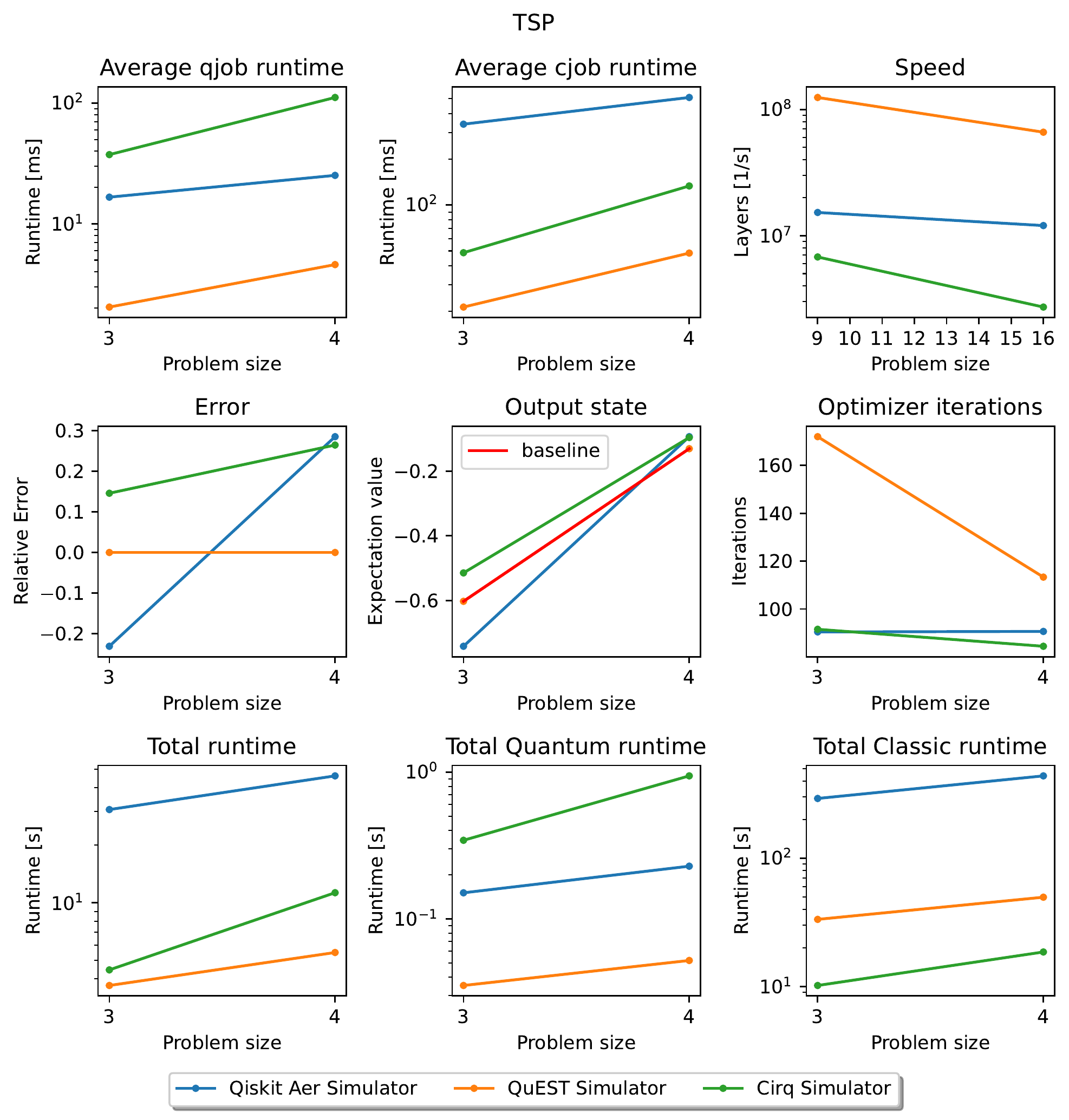}
    \caption{Benchmark results for the Traveling Salesperson Problem}
    \label{fig:app_loc_sim_tsp}
\end{figure}

\begin{figure}[h!]
    \centering
    \includegraphics[width = \mywidthx]{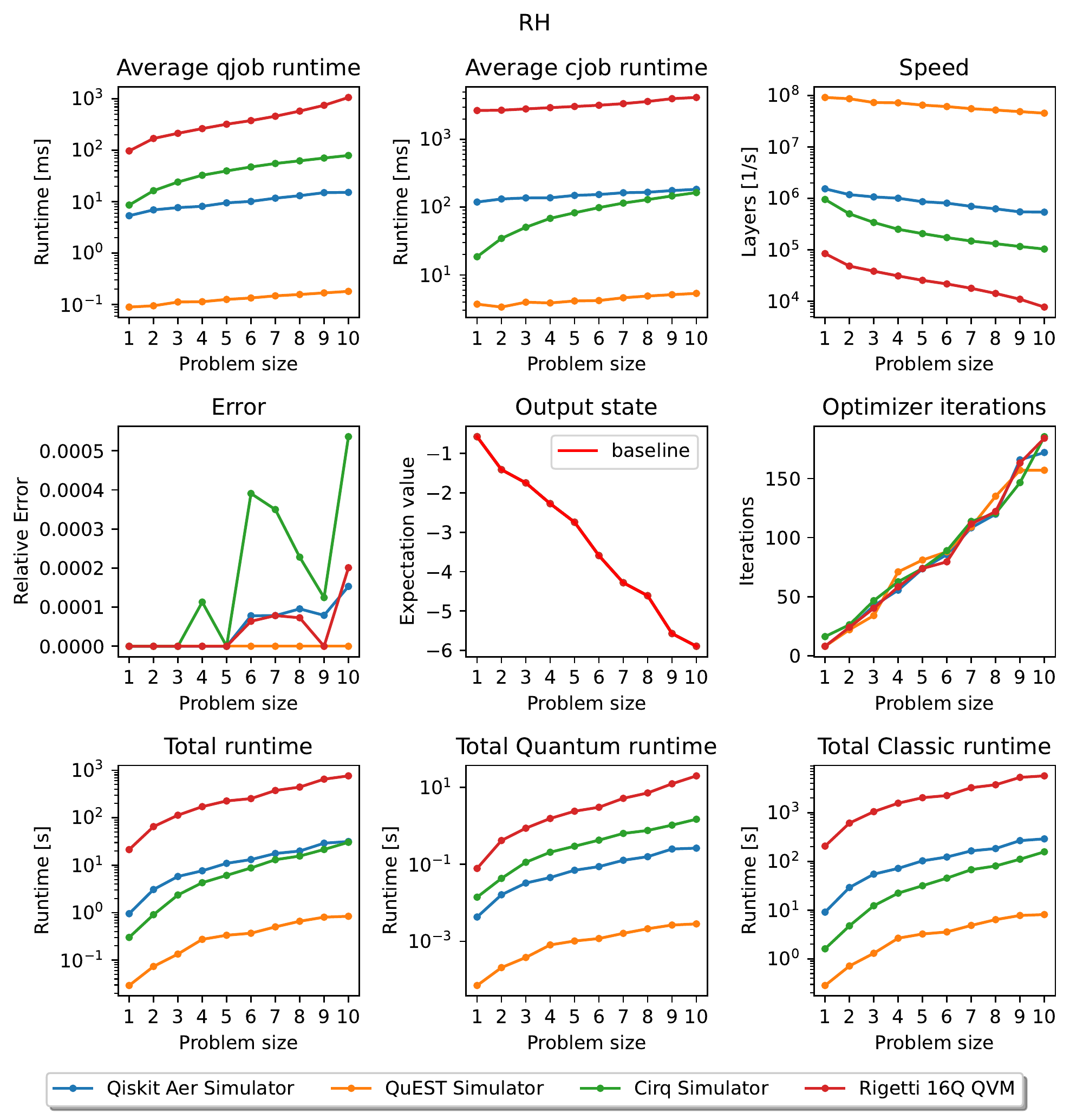}
    \caption{Benchmark results for the Random Diagonal Hamiltonian Problem}
    \label{fig:app_loc_sim_rh}
\end{figure}

\begin{figure}[h!]
    \centering
    \includegraphics[width = \mywidthx]{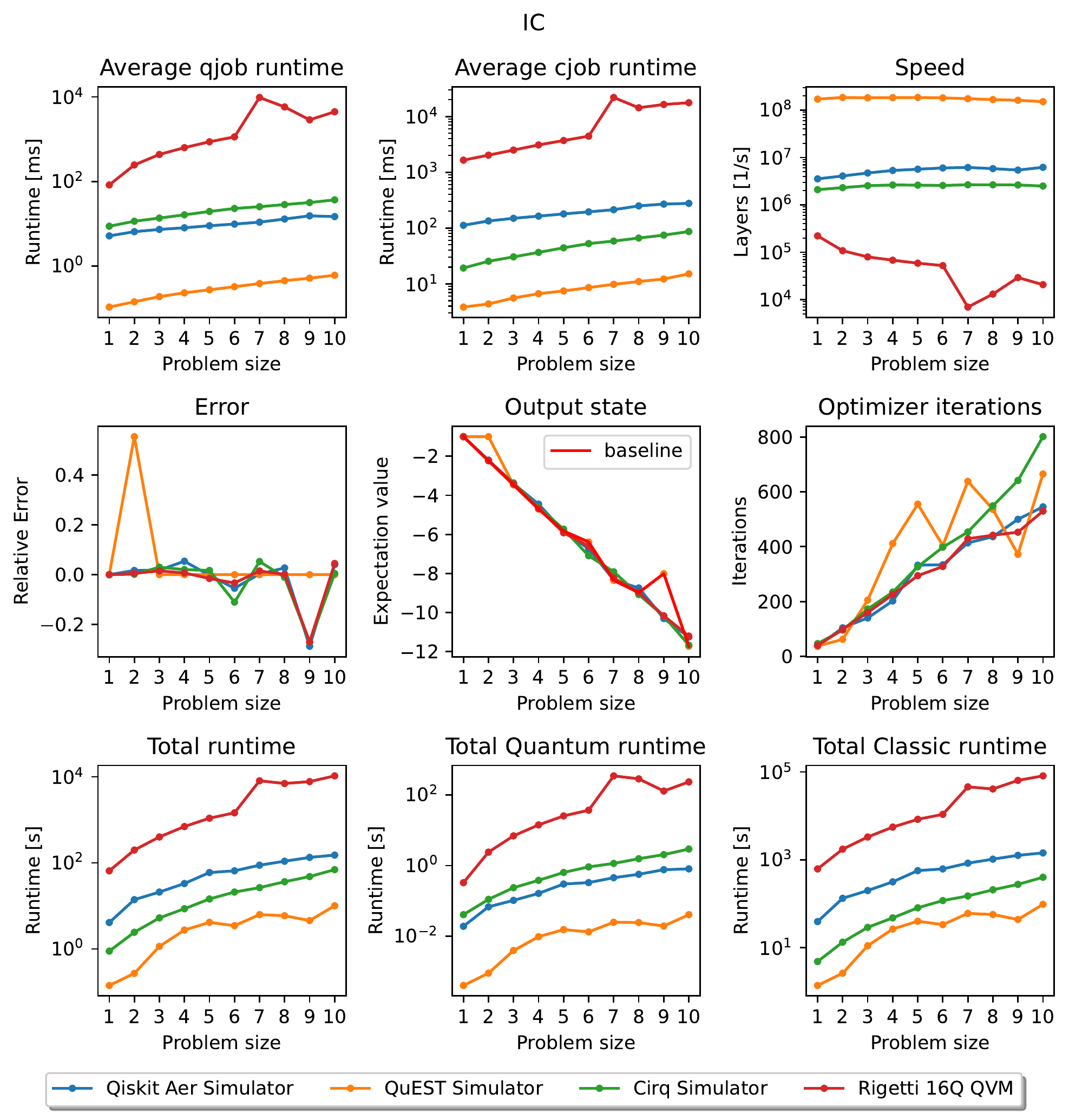}
    \caption{Benchmark results for the Ising Chain model}
    \label{fig:app_loc_sim_ic}
\end{figure}

This subsection lists data of the Qiskit Aer~\cite{qiskit_aer}, Cirq~\cite{cirq_simulator}, Rigetti QVM~\cite{rigetti_qvm} and QuEST~\cite{quest} simulators. Data was collected using a desktop computer, utilizing an AMD Ryzen 5 3600 6-core CPU~\cite{ryzen5} with 16 GB RAM. Data plots are shown in Figures \ref{fig:app_loc_sim_mcp}, \ref{fig:app_loc_sim_dsp}, \ref{fig:app_loc_sim_mis}, \ref{fig:app_loc_sim_tsp}, \ref{fig:app_loc_sim_rh} and \ref{fig:app_loc_sim_ic}.

\leavevmode\thispagestyle{empty}\newpage
\leavevmode\thispagestyle{empty}\newpage
\leavevmode\thispagestyle{empty}\newpage
\subsection{Remote Simulators}
\begin{figure}[h!]
    \centering
    \includegraphics[width = \mywidthx]{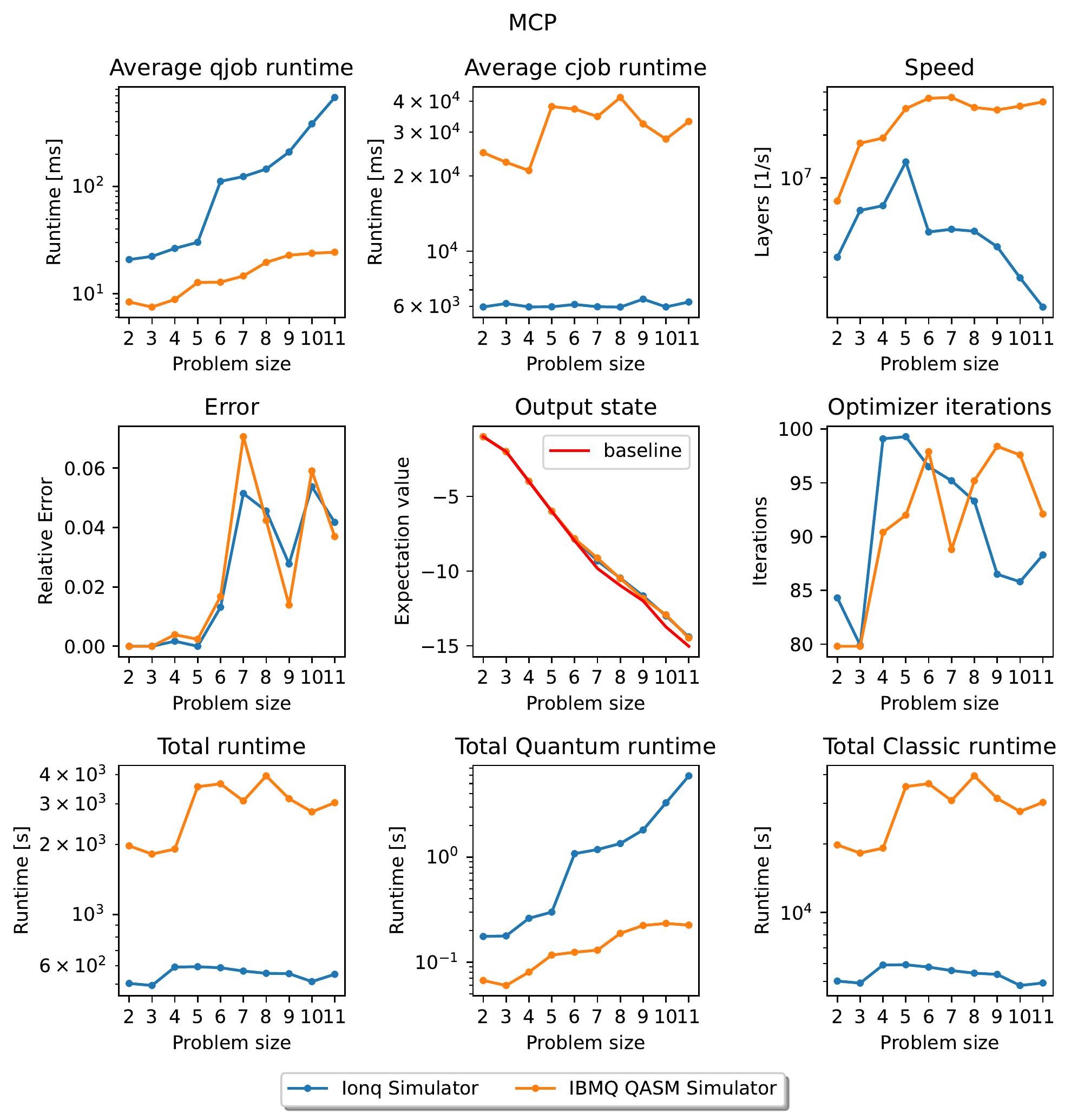}
    \caption{Benchmark results for the MaxCut Problem}
    \label{fig:app_rem_sim_mcp}
\end{figure}

\begin{figure}[h!]
    \centering
    \includegraphics[width = \mywidthx]{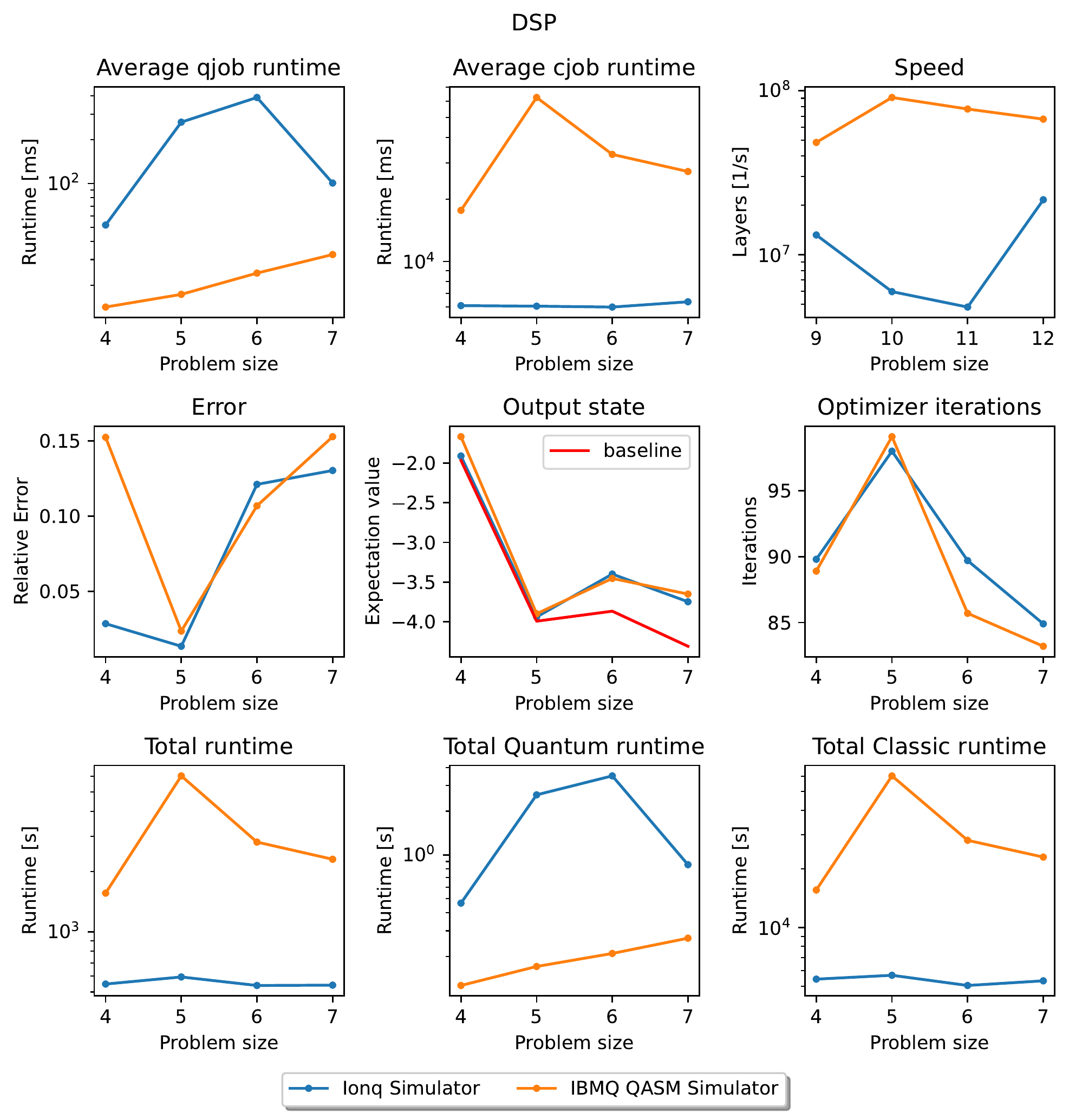}
    \caption{Benchmark results for the Dominating Set Problem}
    \label{fig:app_rem_sim_dsp}
\end{figure}

\begin{figure}[h!]
    \centering
    \includegraphics[width = \mywidthx]{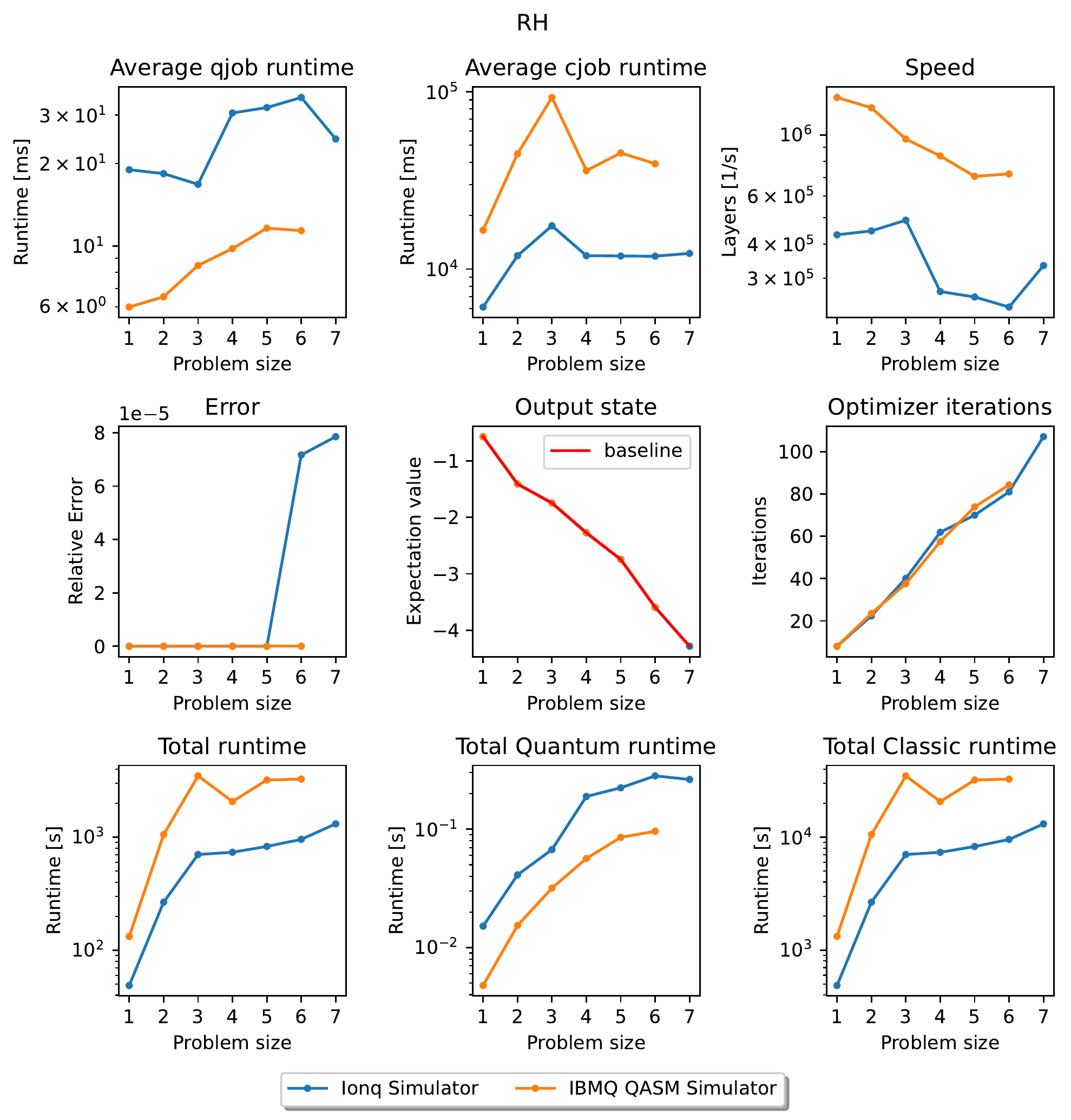}
    \caption{Benchmark results for the Random Diagonal Hamiltonian Problem}
    \label{fig:app_rem_sim_rh}
\end{figure}

\begin{figure}[h!]
    \centering
    \includegraphics[width = \mywidthx]{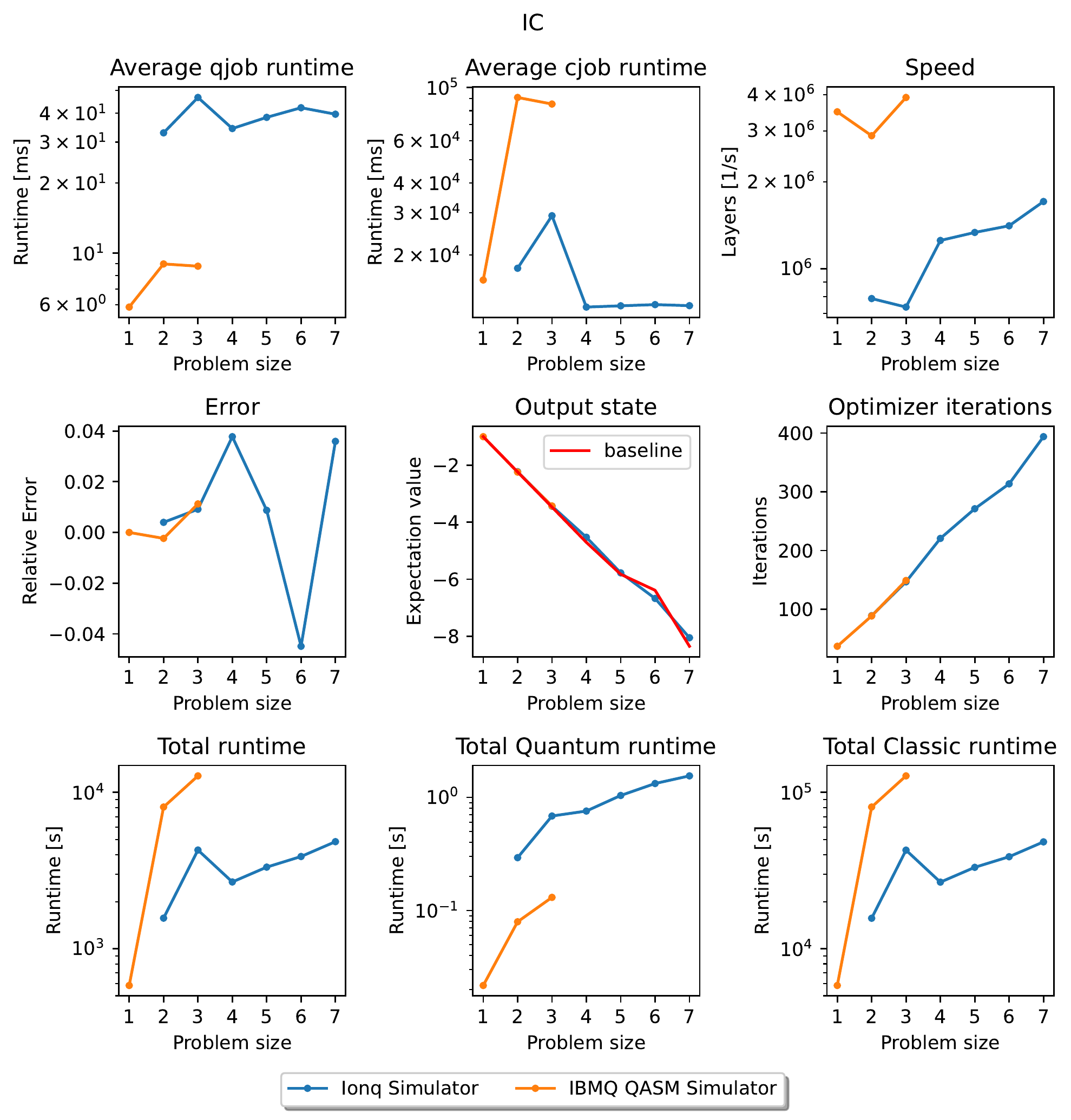}
    \caption{Benchmark results for the Ising Chain model}
    \label{fig:app_rem_sim_ic}
\end{figure}

Data collected from the cloud-accessible  IBMQ QASM Simulator~\cite{ibmq_qasm_simulator} and IonQ Simulator~\cite{ionq_simulator}. Figures \ref{fig:app_rem_sim_mcp}, \ref{fig:app_rem_sim_dsp},  \ref{fig:app_rem_sim_rh} and \ref{fig:app_rem_sim_ic}.

\leavevmode\thispagestyle{empty}\newpage
\leavevmode\thispagestyle{empty}\newpage
\subsection{Remote Hardware}

\begin{figure}[h!]
    \centering
    \includegraphics[width = \mywidthx]{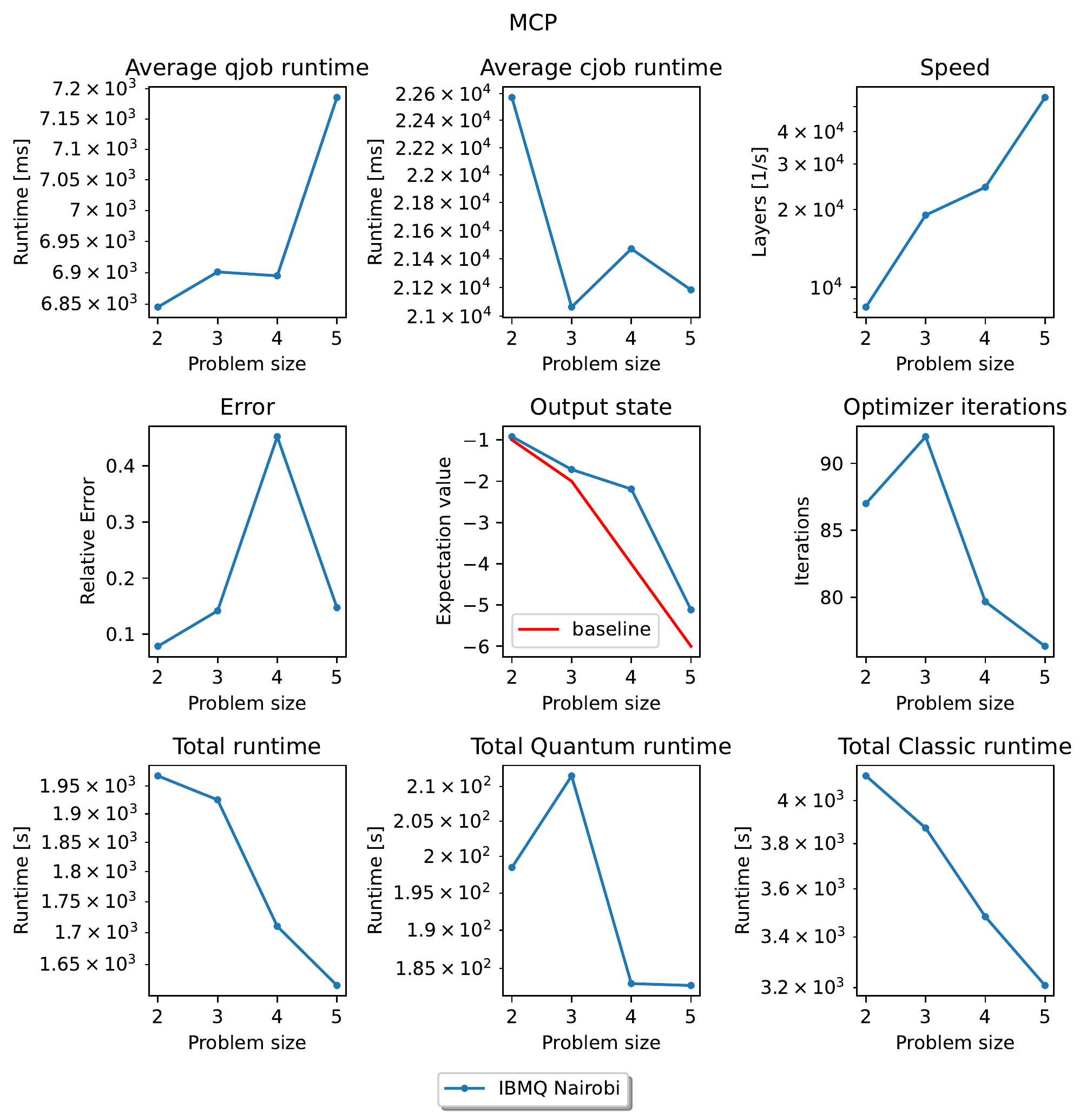}
    \caption{Benchmark results for the MaxCut Problem}
    \label{fig:app_rem_qpu_mcp}
\end{figure}

\begin{figure}[h!]
    \centering
    \includegraphics[width = \mywidthx]{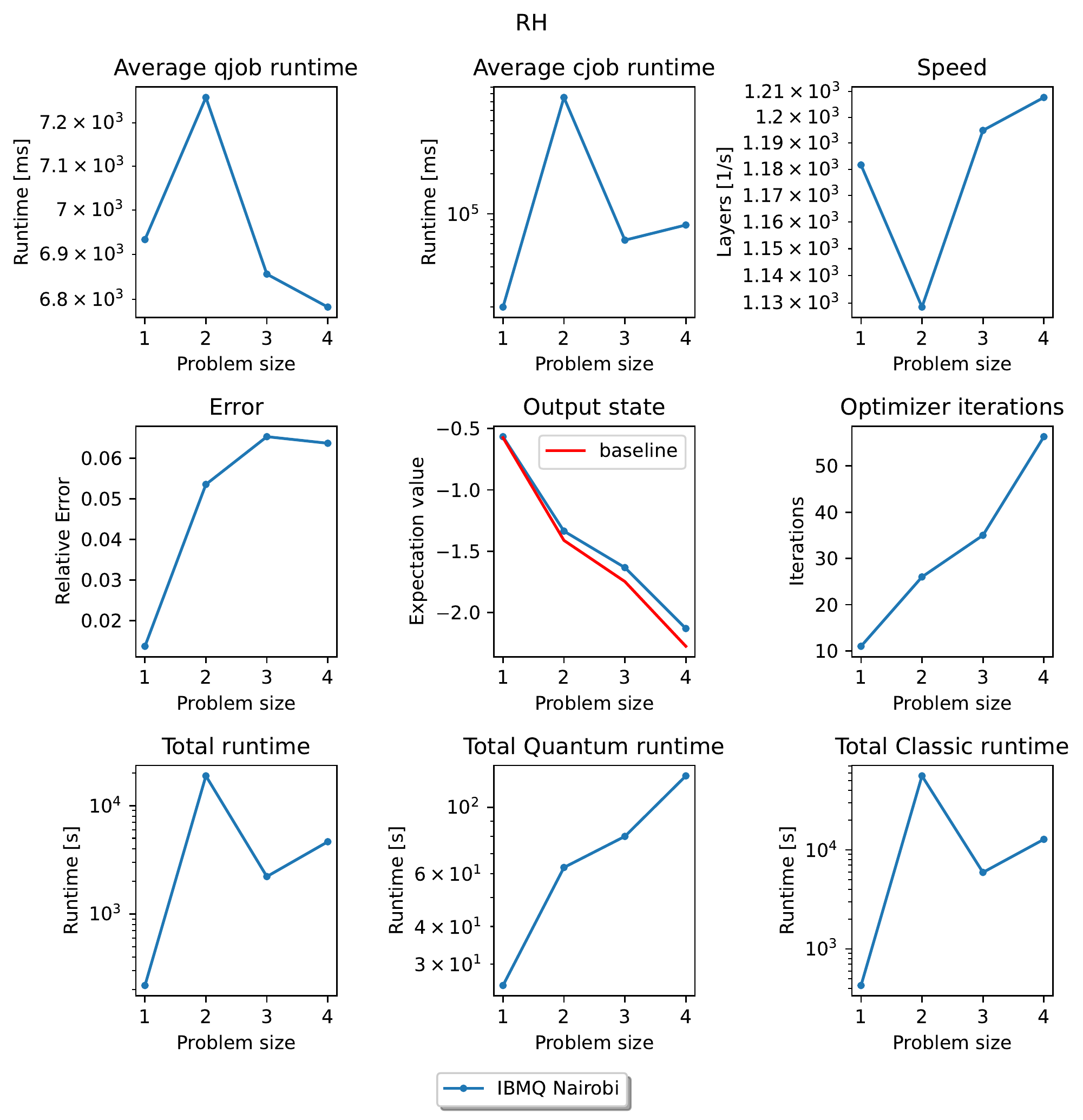}
    \caption{Benchmark results for the Random Diagonal Hamiltonian Problem}
    \label{fig:app_rem_qpu_rh}
\end{figure}

\begin{figure}[h!]
    \centering
    \includegraphics[width = \mywidthx]{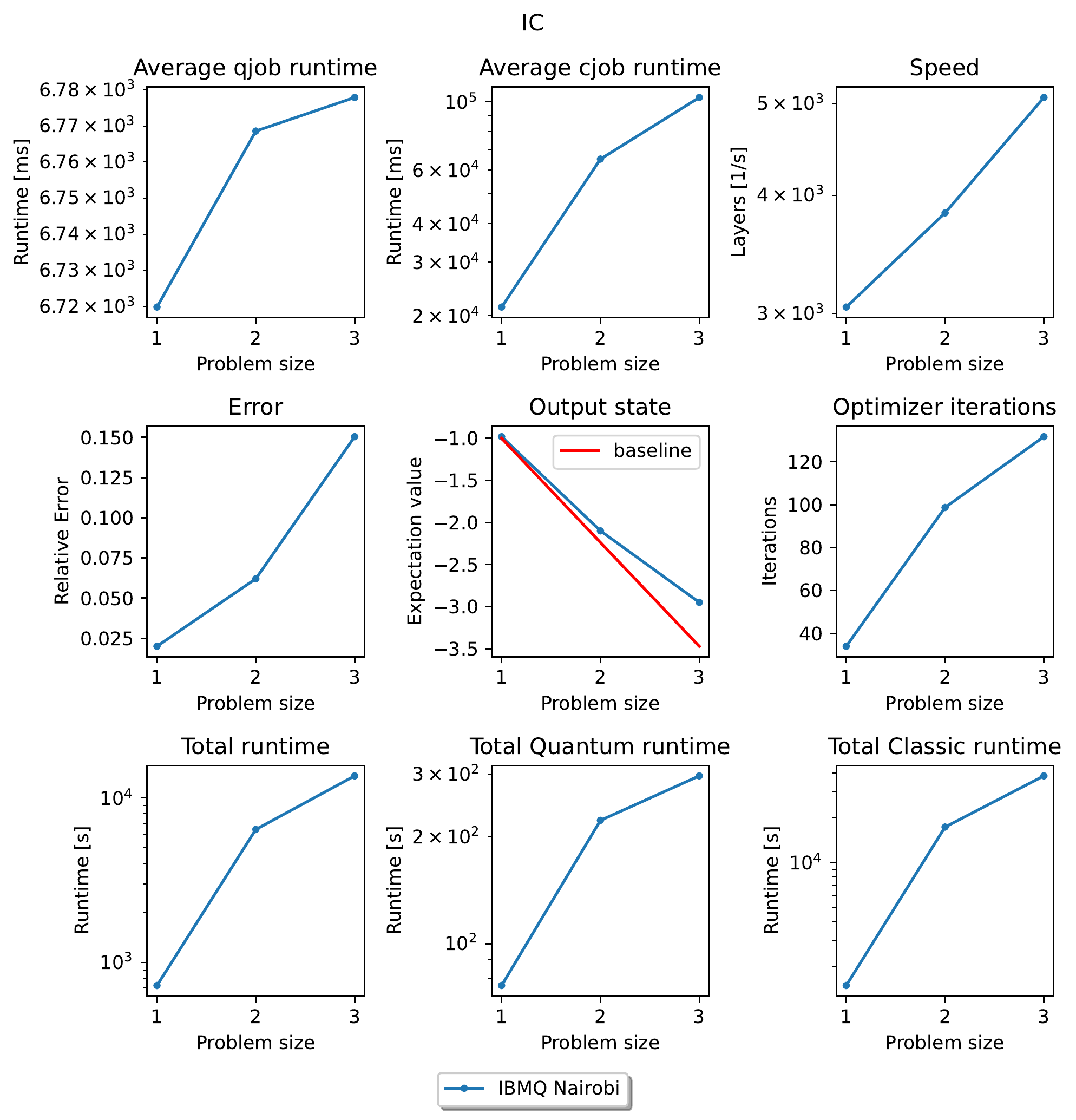}
    \caption{Benchmark results for the Ising Chain model}
    \label{fig:app_rem_qpu_ic}
\end{figure}

Data obtained using the IBMQ Nairobi 7-qubit quantum processor~\cite{ibm_aviary}. Here, only the MaxCut problem, Random Hamiltonian and Ising Chain problem were executed, due to larger qubit requirements for the Dominating Set, Maximum Independent Set and Traveling Salesperson problems. Figures \ref{fig:app_rem_qpu_mcp},  \ref{fig:app_rem_qpu_rh} and \ref{fig:app_rem_qpu_ic}.

\leavevmode\thispagestyle{empty}\newpage
\leavevmode\thispagestyle{empty}\newpage
\section{Complete Benchmark Results} \label{app:bench_result}
This section lists all benchmark results per tested devices of which data was collected and presented in Appendix \ref{app:datasets}. The figures show the sub-scores per problem set, as well as the overall benchmark score per quantum device.

\newcommand\mywidth{0.89\linewidth}

\subsection{Local Simulators}
\begin{figure}[h!]
    \centering
    \includegraphics[width = \mywidth]{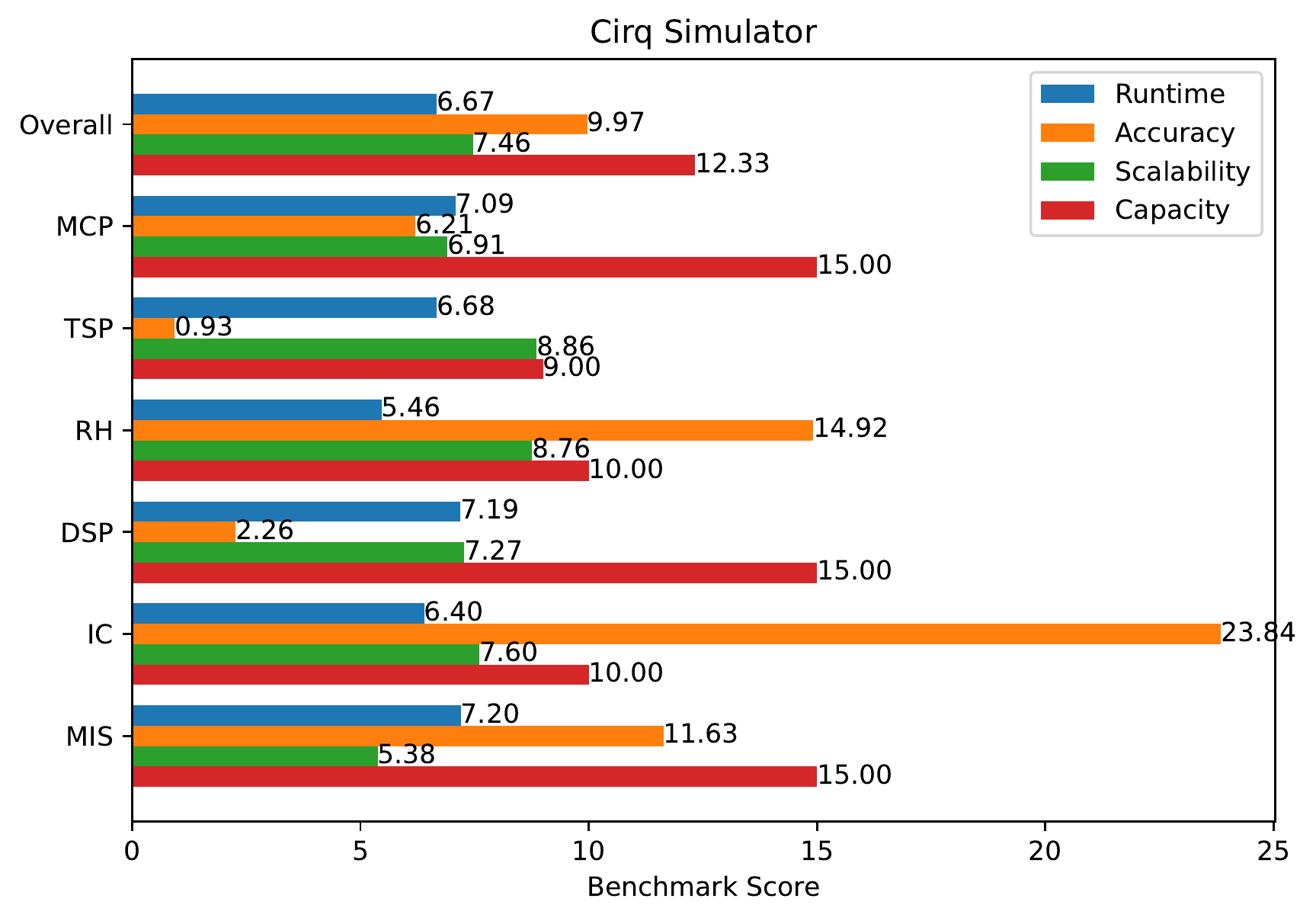}
    \caption{Benchmark results for the localCirq simulator}
    \label{fig:app_cirq_sim}
\end{figure}

\begin{figure}[h!]
    \centering
    \includegraphics[width = \mywidth]{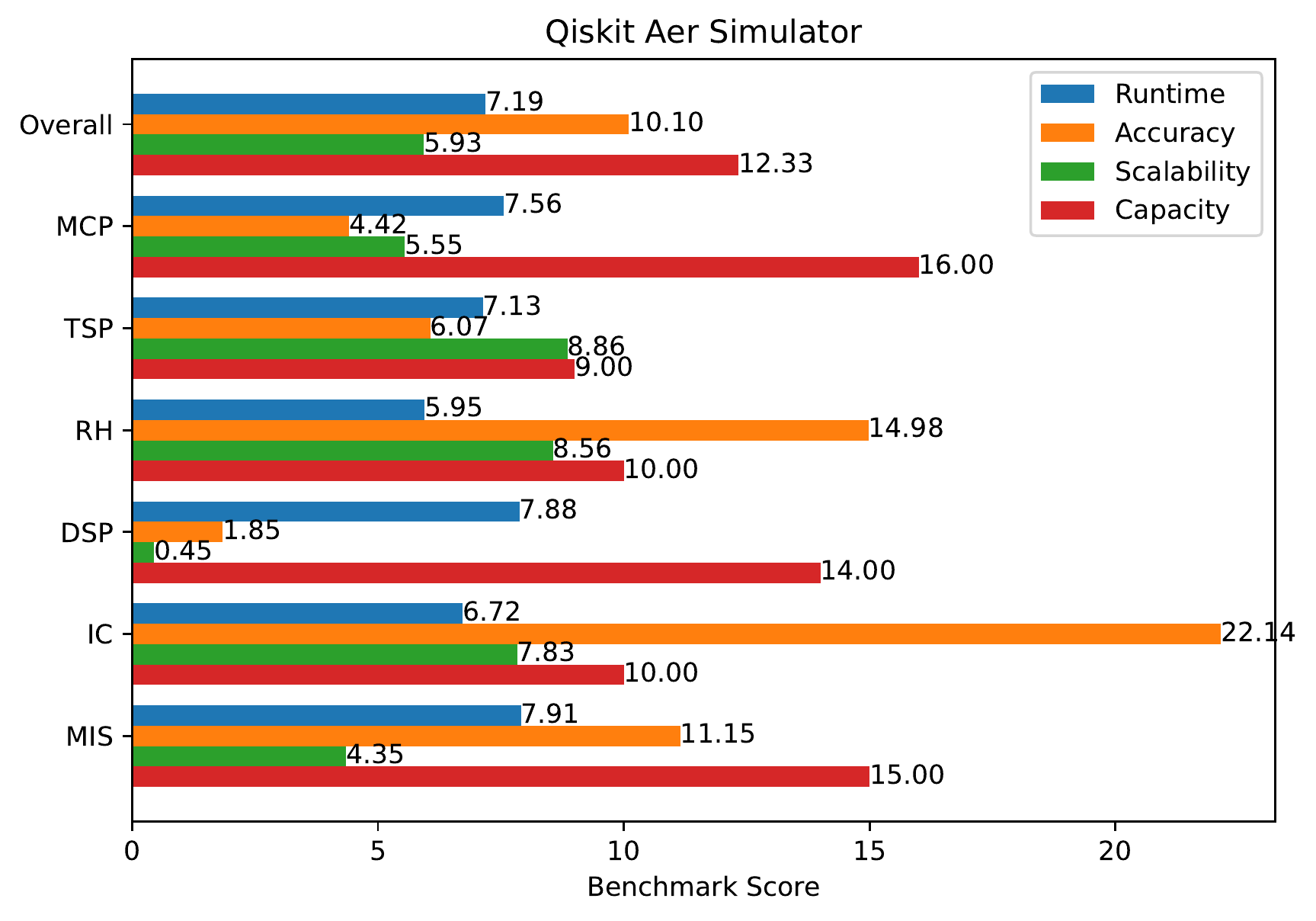}
    \caption{Benchmark results for the local Qiskit Aer simulator}
    \label{fig:app_qiskit_aer_sim}
\end{figure}

\begin{figure}[h!]
    \centering
    \includegraphics[width = \mywidth]{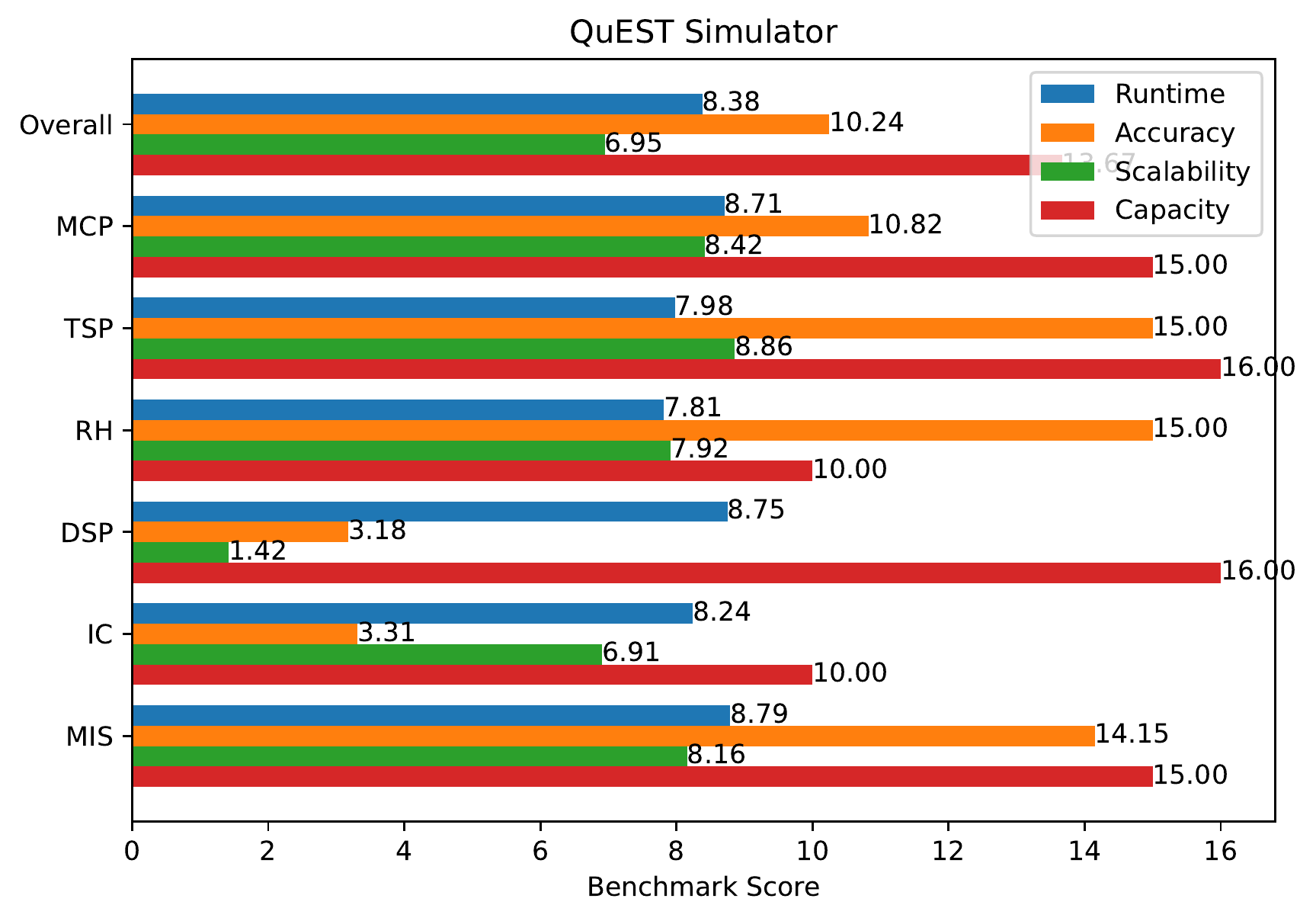}
    \caption{Benchmark results for the local QuEST simulator}
    \label{fig:app_quest_sim}
\end{figure}

\begin{figure}[h!]
    \centering
    \includegraphics[width = \mywidth]{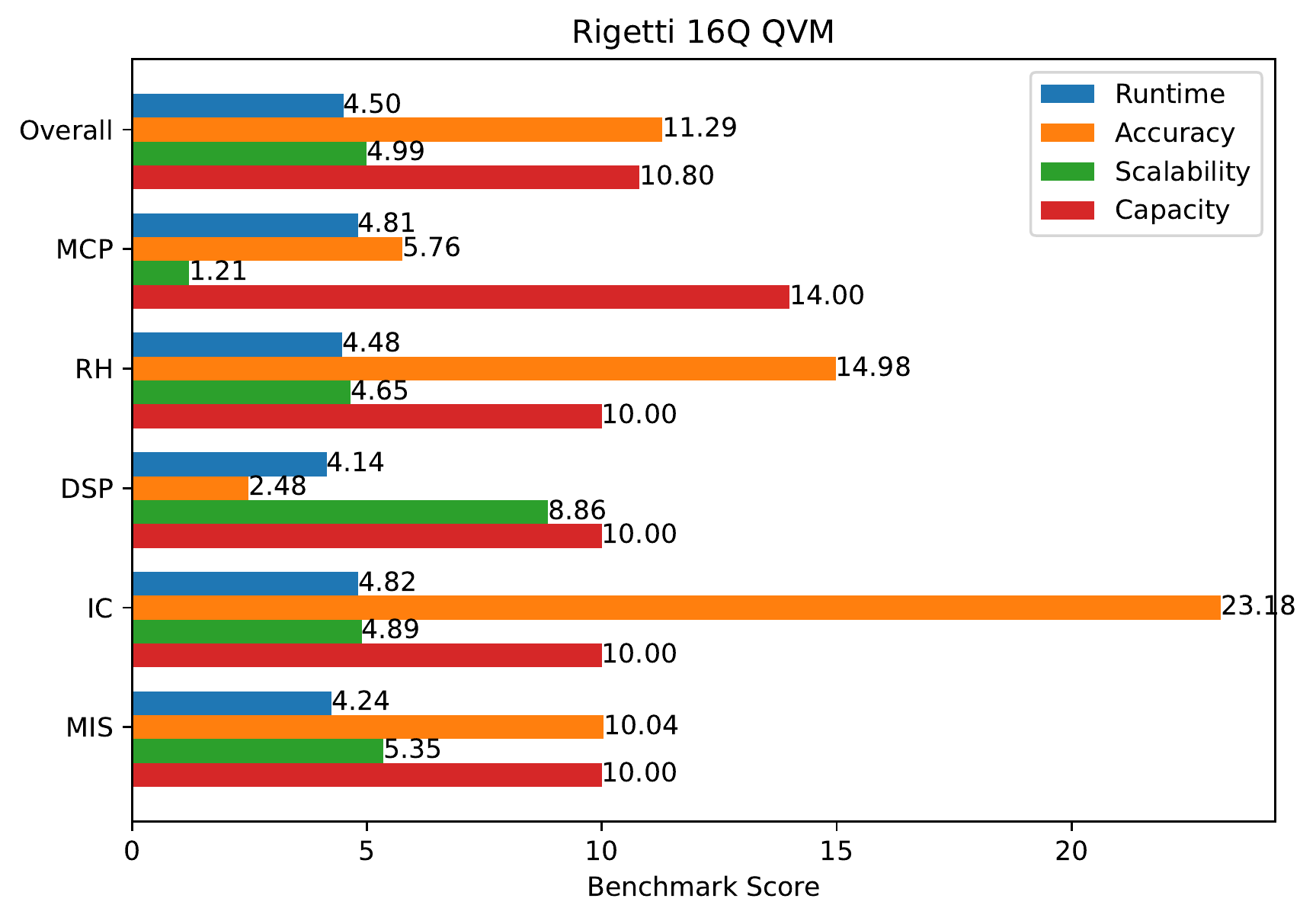}
    \caption{Benchmark results for the local Rigetti 16Q QVM simulator}
    \label{fig:app_qvm_sim}
\end{figure}

Benchmark results for the Cirq~\cite{cirq_simulator}, Qiskit Aer~\cite{qiskit_aer}, QuEST~\cite{quest} and Rigetti QVM~\cite{rigetti_qvm} simulators, shown in Figures \ref{fig:app_cirq_sim}, \ref{fig:app_qiskit_aer_sim}, \ref{fig:app_quest_sim} and \ref{fig:app_qvm_sim} respectively.

\leavevmode\thispagestyle{empty}\newpage
\subsection{Remote Simulators}
\begin{figure}[h!]
    \centering
    \includegraphics[width = \mywidth]{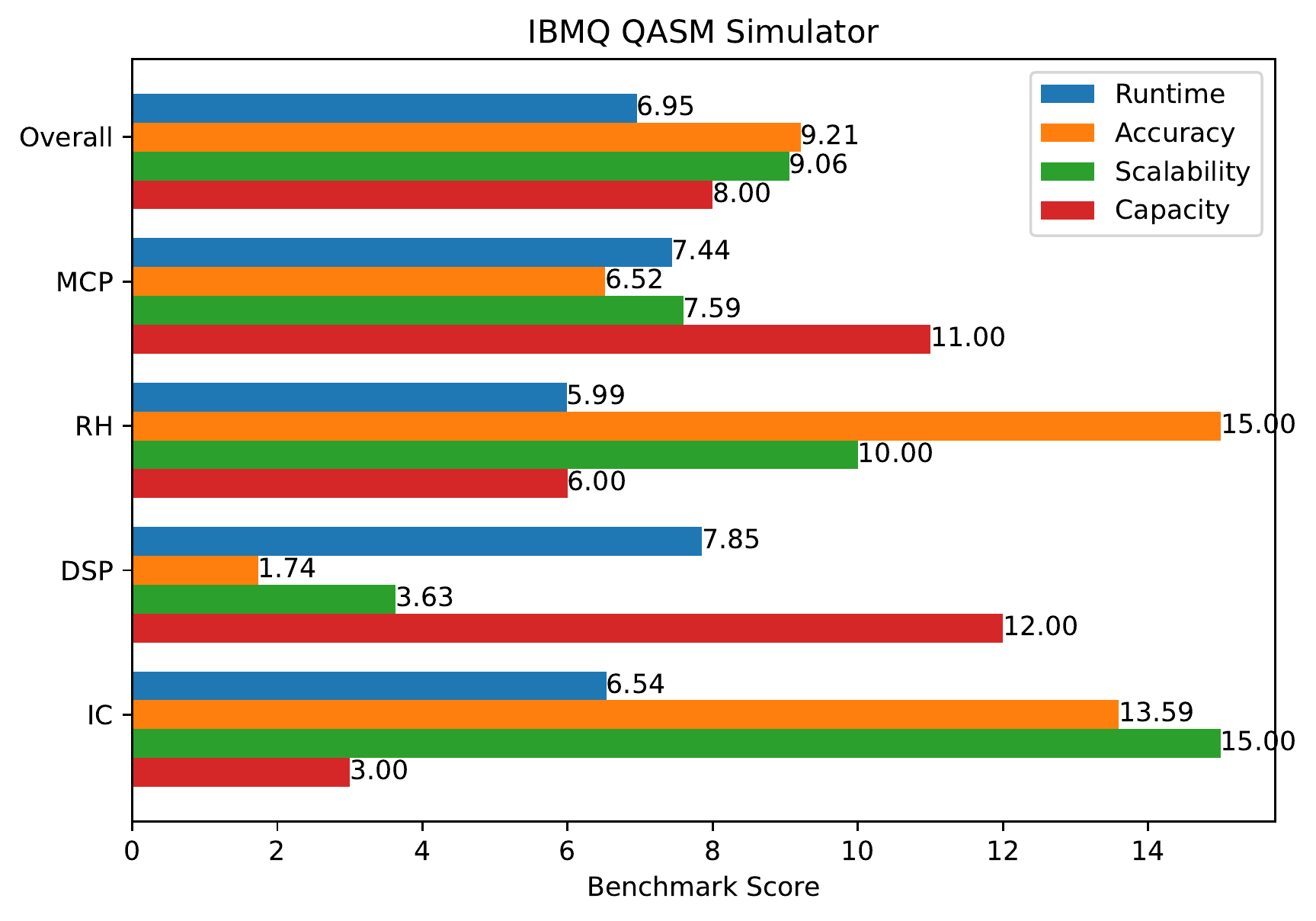}
    \caption{Benchmark results for the remote IBMQ QASM simulator}
    \label{fig:app_ibmq_qasm_sim}
\end{figure}

\begin{figure}[h!]
    \centering
    \includegraphics[width = \mywidth]{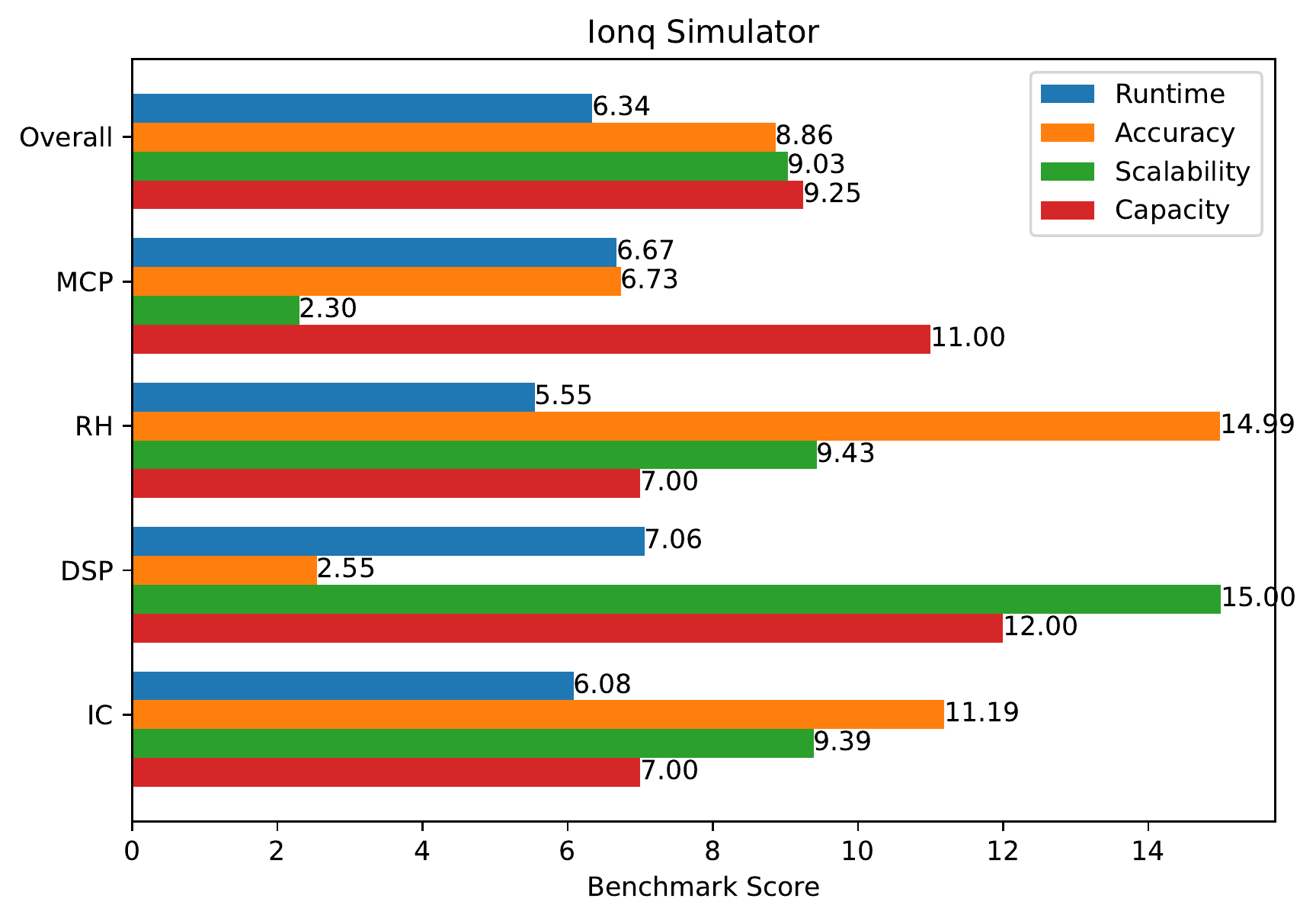}
    \caption{Benchmark results for the remote IonQ simulator}
    \label{fig:app_ionq_sim}
\end{figure}

Benchmark results for the cloud-accessible IBMQ QASM Simulator~\cite{ibmq_qasm_simulator} and IonQ Simulator~\cite{ionq_simulator}. Figures  \ref{fig:app_ibmq_qasm_sim} and \ref{fig:app_ionq_sim} respectively show their individual benchmark results.

\subsection{Remote Hardware}
\begin{figure}[h!]
    \centering
    \includegraphics[width = \mywidth]{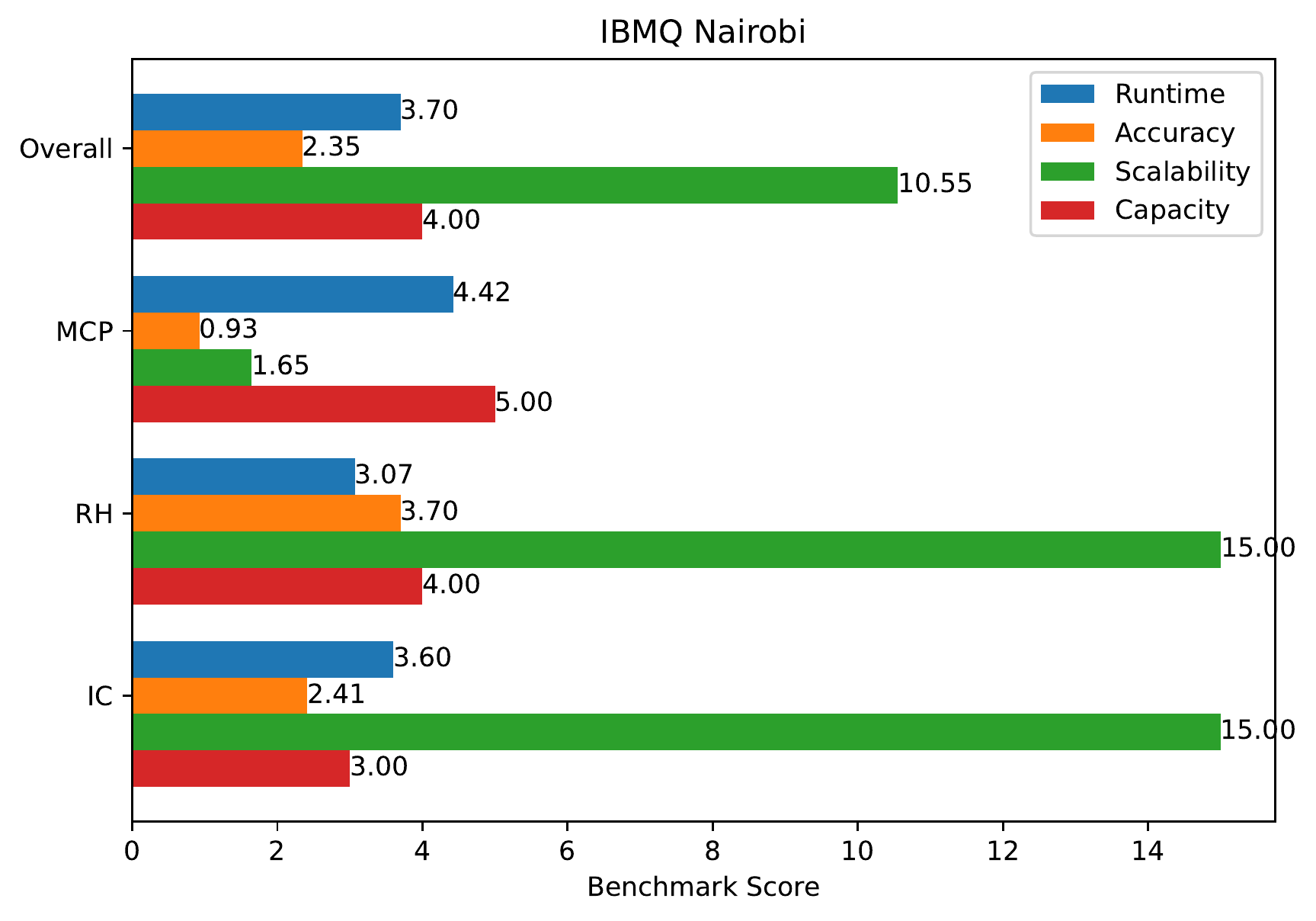}
    \caption{Benchmark results for the remote IBMQ Nairobi quantum processor}
    \label{fig:app_ibmq_nairobi}
\end{figure}

Benchmark result for the IBMQ Nairobi quantum processor~\cite{ibm_aviary}, Figure \ref{fig:app_ibmq_nairobi}.

\end{appendices}

\end{document}